\documentclass{aa}

\usepackage{graphicx}
\usepackage{xcolor}

\usepackage{txfonts}
\usepackage{hyperref}
\usepackage[version=4]{mhchem}
\usepackage{lscape}
\usepackage{arydshln}
\usepackage{subcaption} 
\usepackage{float}

\begin{document} 
\title{Mid-infrared extinction curve for protostellar envelopes from JWST-detected embedded jet emission: the case of TMC1A}
   \subtitle{}

   \author{K. D. Assani \inst{1,2}
          \and Z.-Y. Li \inst{1, 2}
          \and J. P. Ramsey \inst{3}
          \and Ł. Tychoniec \inst{4, 5}
          \and L. Francis \inst{4}
          \and V. J. M. Le Gouellec \inst{6,7}
          \and A. Caratti o Garatti \inst{8}
          \and T. Giannini \inst{8}
          \and M. McClure \inst{4}
          \and P. Bjerkeli \inst{9}
          \and H. Calcutt \inst{9}
          \and H. Beuther \inst{10}
          \and R. Devaraj \inst{11}
          \and X. Liu \inst{5}
          \and A. Plunkett \inst{12}
          \and M. G. Navarro \inst{6}
          \and E. F. van Dishoeck \inst{4}
          \and D. Harsono\inst{13}
          }

   \institute{
   Department of Astronomy, University of  Virginia, Charlottesville, VA 22903, USA 
    \and
    Virginia Institute of Theoretical Astronomy, University of Virginia, Charlottesville, VA 22903, USA 
   \and
   Bluedrop Training \& Simulation, Inc., 36 Solutions Drive \#300, Halifax, Nova Scotia B3S 1N2, Canada 
   \and 
    Leiden Observatory, Leiden University, PO Box 9513, 2300RA, Leiden, The Netherlands  
   \and 
   European Southern Observatory, Karl-Schwarzschild-Strasse 2, 85748 Garching bei München, Germany 
   \and 
   Institut de Ciències de l’Espai (ICE-CSIC), Campus UAB, Can Magrans S/N, E-08193 Cerdanyola del Vallès, Catalonia, Spain  
   \and 
    Institut d’Estudis Espacials de Catalunya (IEEC), c/Gran Capita, 2-4, E-08034 Barcelona, Catalonia, Spain 
   \and 
   INAF-Osservatorio Astronomico di Capodimonte, Salita Moiariello 16, 80131 Napoli, Italy  
   \and
   Chalmers University of Technology, Department of Space, Earth and Environment, 412 96 Gothenburg, Sweden 
   \and
   Max-Planck-Institut für Astronomie, Königstuhl 17, 69117 Heidelberg, Germany 
   \and 
   Dublin Institute for Advanced Studies, DIAS Headquarters, 10 Burlington Road, D04C932 Dublin, Ireland 
   \and 
   National Radio Astronomy Observatory, 520 Edgemont Road, Charlottesville, VA 22903, USA 
   \and
   Institute of Astronomy, Department of Physics, National Tsing Hua University, Hsinchu, Taiwan 
   }

 
  \abstract
    {Dust grains are fundamental components of the interstellar medium and play a crucial role in star formation, serving as catalysts for chemical reactions and the building blocks of planets. Extinction curves serve as a tool for characterizing dust properties, yet mid-infrared (MIR) extinction remains less constrained in protostellar environments. Gas-phase line ratios from embedded protostellar jets offer a spatially resolved method to measure extinction from protostellar envelopes, complementing traditional background starlight techniques.}
    {We aim to derive mid-IR extinction curves along the lines of sight toward a protostellar jet embedded within an envelope and assess whether they differ from those inferred in dense molecular clouds.}
    {We analyze JWST NIRSpec IFU and MIRI MRS observations, focusing on four locations along the blue-shifted TMC1A jet. After extracting observed [\ion{Fe}{II}] line intensities, we model intrinsic line ratios using the Cloudy spectral synthesis code across a range of electron densities and temperatures. By comparing observed near-IR (NIR) and MIR line ratios to Cloudy-predicted intrinsic ratios, we infer the relative extinction between NIR and MIR wavelengths.}
    {Electron densities ($n_e$) derived from NIR [\ion{Fe}{II}] lines range from $\sim5 \times 10^4$ to $\sim5 \times 10^3$ cm$^{-3}$ along the jet axis at scales $\lesssim$350 AU, serving as reference points for comparing relative NIR and MIR extinction. The derived MIR extinction values exhibit higher reddening than the empirical dark cloud curve from \citet{mcclure_extinction_law_2009} at the corresponding $n_e$ and temperatures ranging from a few 10$^3$ K to $\sim$10$^4$ K, as adopted from shock models. While both electron density and temperature influence the NIR-to-MIR [\ion{Fe}{II}] line ratios, the ratios are more strongly dependent on $n_e$ over the adopted range. If the MIR emission originates from gas that is less dense and cooler than the NIR-emitting region, the inferred extinction curves remain consistent with background star-derived values.}
    {This study introduces a new line-based method for deriving spatially resolved MIR extinction curves towards embedded protostellar sources exhibiting a bright [\ion{Fe}{II}] jet. The results suggest that protostellar envelopes may contain dust with a modified grain size distribution---such as an increased fraction of larger grains, potentially due to grain growth---if the MIR and NIR lines originate from similar regions along the same sightlines. Alternatively, if the grain size distribution has not changed (i.e., no grain growth), the MIR lines may trace cooler, less dense gas than the NIR lines along the same sightlines. This method provides a novel approach for studying dust properties in star-forming regions and can be extended to other protostellar systems to refine extinction models in embedded environments.}

   \titlerunning{Spatially resolved extinction curves toward TMC1A with JWST}
   \maketitle

\section{Introduction} \label{Sec: introduction}

Despite the dust-to-gas mass ratio of 0.01, dust grains are fundamental components of the interstellar medium (ISM), produced during the final stages of stellar evolution and subsequently processed in interstellar environments. These grains act as catalysts for chemical reactions, enabling the formation of complex organic molecules and facilitating key processes essential to star and planet formation \citep[e.g.,][]{Johansen_planetesimal_formation_2014,brugger_pebbles_vs_planetesimals_2020,Johansen_pebble_accretion_terrestrial_2021,Boogert_2015_icy_universe,McClure_ice_age_nature_2023}.  The earliest stages of planet formation begin with the growth of dust grains, a process that appears to commence early in the star formation sequence \citep[e.g.,][]{Drazkowska_planet_formation_ALMA_Kepler_era_2024}. 
While JWST is beginning to unravel dust grain properties through infrared observations along various lines of sight within molecular clouds \citep[e.g.,][]{Decleir_MEAD_2025}, it remains less clear how these dust properties persist as a protostar forms and creates a dense environment within a collapsing cloud.

At millimeter wavelengths, low spectral index values---referring to the shallow slope of flux density across the mm regime---are often interpreted as evidence for grain growth, suggesting the presence of large grains in the disks and infalling envelopes of the youngest protostars (ages $\sim$10$^{4-5}$ years) \citep[e.g.,][]{Miotello_A_classI_protostars_mm_sized_2014, Galametz_Maury_class0_grain_growth_2019, Cacciapuoti_classI_grain_properties_infal_rates_2024}. While the sizes of dust grains contributing to low dust emissivities remain less well constrained, models of dust polarization based on radiative alignment torques (RATs) require grains larger than $\sim$10 $\mu$m to explain observed polarization levels in protostellar envelopes at millimeter wavelengths \citep{Valdivia_polarization_grain_growth_2019, Le_Gouellec_dust_grains_polarization_serpens_2019}. 

Across infrared (IR) to ultraviolet (UV) wavelengths, dust grain properties are typically inferred from extinction curves and/or solid-state features of the refractories and/or volatiles. Extinction curves describe how dust grains absorb and scatter light as a function of wavelength, providing a critical tool for probing their composition, size distribution, and processing in astrophysical environments. Observationally, extinction curves are usually derived by measuring the attenuation of stellar photospheric continuum emission along different lines of sight, revealing how dust interacts with radiation in various astrophysical environments \citep[e.g.,][]{cardelli1989relationship, mcclure_extinction_law_2009, gordon_extinction_2023}. Theoretically, extinction models connect these observational trends to underlying dust grain populations by adjusting grain size distributions and compositions to match the observed extinction laws \citep[e.g.,][]{Weingartner_and_Draine_01, Craspi_firstKP5_2008, pontoppidan_extinction_2024}. 

\citet{cardelli1989relationship} introduced a widely adopted UV to near-infrared (NIR) extinction law derived from line-of-sight observations of bright stars with known spectral types, showing that the shape of the extinction curve depends on the total-to-selective extinction ratio (R$_V = A_V / E(B-V)$), where A$_V$ is the total extinction in the V-band and $E(B-V)$ represents the difference in extinction between the B and V bands. Notably, the observed extinction curves revealed that extinction in the NIR ($\lambda\sim 1-5 $ $\mu$m) appears independent of R$_V$ while a larger dependence is found in UV (the "Far-UV rise").  In the mid-infrared (MIR), \citet{mcclure_extinction_law_2009} analyzed 5-20 $\mu$m Spitzer IRS spectra of background G0–M4 III stars and confirmed that extinction curves derived from stars behind dense dark clouds exhibit higher levels of mid-IR extinction compared to those behind less opaque clouds, which resemble the diffuse interstellar medium (ISM). The higher extinction regimes correspond to flatter MIR extinction curves relative to the NIR that aligns closely with the R$_V=5.5$, case B, curve in \cite{Weingartner_and_Draine_01}. 

From a modeling perspective, the observed extinction differences relative to optical/IR across lines of sight in the UV ($\lambda < 0.16\mu$m) and the mid-infrared (5–30 $\mu$m) correspond to differences in dust composition, grain size distributions, and the presence of solid-state features such as refractories and volatiles. The R$_V$ = 5.5 curve \citep[Case B in][]{Weingartner_and_Draine_01} shows higher extinction at $\lambda>2$ $\mu$m compared to the R$_V$ = 3.1 case, attributed to a decrease in small grains (<0.1 $\mu$m) and an increased fraction of larger silicate and carbonaceous grains, resulting in the observed flattening of the extinction curve in the MIR relative to the NIR. This flattening effect is also consistent with the "KP5" extinction curve, which models extinction as a function of wavelength for dense clouds and protostellar envelopes, including the effects from ices (e.g., H$_2$O, CO, and CO$_2$) that introduce sharp opacity features at specific mid-infrared wavelengths \citep{Craspi_firstKP5_2008, pontoppidan_extinction_2024}. The KP5 model assumes a grain size distribution predominantly larger than 0.1 $\mu$m, with an upper size limit of 7 $\mu$m, consistent with the grain size distribution from \cite{Weingartner_and_Draine_01} and broadly consistent with that measured for two lines of sight within a dense molecular cloud by \cite{Dartois_spectroscopic_sizing_ICE_AGE_2024} using radiative transfer modeling of the scattering wings of ice features towards two background stars in Chameleon I.

With the empirically derived MIR extinction curve from \cite{mcclure_extinction_law_2009} showing that denser lines of sight ($A_K > 1$) align more closely with the R$_V=5.5$ Case B curve while more diffuse sightlines ($A_K < 1$) resemble the R$_V=3.1$ curve, there is growing evidence that dust grains grow larger in denser regions. Extending this trend, should we expect an even flatter MIR extinction curve in the denser regions ($A_K \gtrsim 2$) surrounding a protostar---such as its envelope or disk---where grain size distributions (e.g., due to grain growth) may differ from those inferred in molecular clouds?

One way to probe MIR extinction toward embedded environments is through protostellar outflows, which emit strong atomic and molecular lines that become attenuated by the surrounding dust. Unlike background starlight methods, which require bright sources behind dusty regions, gas emission lines provide a spatially resolved approach to measuring extinction within the envelope itself. These outflows exhibit a hierarchical arrangement of nested structures, commonly traced by species such as [\ion{Fe}{II}], H$_2$, and CO, distinguishing between collimated jets, wider molecular outflows, and intermediate-angle disk winds \citep[e.g.,][]{harsono2023_tmc1a, pascucci2024nested, JOYS_Tychoniec_TMC1_2024, Dougados_DG_Tau_B_nestingH2_CO_2024, JOYS_Garatti_HH211_textbook_2024, Le_Gouellec_embedded_NIR_jet_S68N_2024}.

Gas emission lines from protostellar jets are produced from radiative shocks, where supersonic outflows collide with more slowly moving material, creating sharp transitions at the shock front \citep[e.g.,][]{franketal14_ppvi}. At this interface, gas is rapidly compressed and heated, initiating a post-shock cooling zone where atomic and molecular species become emissive \citep[e.g.,][]{hollenbach1989molecule}. For the NIR and MIR [\ion{Fe}{II}] lines, shock models predict that these lines become emissive at electron densities $\sim$10$^3$–10$^5$ cm$^{-3}$ and temperatures $\gtrsim$7,000 K \citep{Nisini_lecture_notes_2008, giannini2015_HH1, koo_2016_shock_models}. It is within this cooling region that atomic and singly-ionized species, including forbidden line emission from Fe (i.e., [\ion{Fe}{II}]), are collisionally excited. These conditions allow [\ion{Fe}{II}] emission to remain strong without being collisionally quenched, making it a reliable tracer of the physical and chemical properties of protostellar jets.

For this study, we focus on the Class I protostar TMC1A (IRAS 04365+2535), a $\sim$0.4 $M_{\odot}$, $\sim$2.7 $L_\odot$ source located in the Taurus Molecular Cloud at a distance of 140 pc \citep{galli2019structure} and disk viewed at an inclination of $\sim$54$^\circ$, representing an evolutionary stage approximately 10$^5$ years after the onset of star formation \citep{kristensen12,harsono18}. Its nested outflow structure consists of a central collimated atomic [Fe II] jet, a wider-angle H$_2$ shell observed with JWST \citep{harsono2023_tmc1a, assani_tmc1a_asymmetry}, and an even broader CO cavity observed with ALMA \citep{aso15, bjerkeli2016, aso2021}. This outflow is embedded within a dense envelope of gas and dust \citep{SCUBA_DI_francesco_2008}, which dominates the extinction along the observer's line of sight. Previous NIR extinction measurements toward TMC1A show that extinction decreases with increasing distance from the protostar, reflecting changes in extinction magnitude (column density) along different sightlines \citep{assani_tmc1a_asymmetry} also seen in other sources \citep{Erkaletal2021, narang2023investigating, Delabrosse_JWST_DGtauB_2024, Valentin_accretion_class0_2024}. 

To compare relative NIR/MIR extinction along the jet as a probe of changes in the dust grain size distribution (i.e., dust growth) within the envelope, we adopt an approach that combines excitation models with multi-wavelength embedded [\ion{Fe}{II}] jet emission line data. We focus on line ratios rather than absolute intensities, as ratios are less sensitive to uncertainties in the emitting surface and source geometry. Emission lines originating from the same upper energy level have traditionally served as robust tools for measuring extinction magnitude, as their intrinsic line ratios depend solely on radiative de-excitation rates, determined by the Einstein A-coefficients, and are independent of physical conditions such as density and temperature \citep[e.g.,][]{Nisini_lecture_notes_2008,giannini2015empirical}. Extending this analysis to include emission lines from different upper energy levels, we leverage excitation models to predict a range of intrinsic line ratios across a broader wavelength span. 

This approach enables us to compute extinction curves from observational line emission data, compare them with the existing extinction laws, and identify potential variations in extinction along different lines of sight. Using JWST NIRSpec IFU and MIRI MRS spectroscopy, we aim to characterize the extinction toward the TMC1A protostellar jet and assess how it compares to extinction laws derived from diffuse and dense cloud environments away from protostars. This study provides a spatially resolved probe of dust properties in a protostellar envelope, offering new insights into how dust evolves during early star formation.

The paper is organized as follows. In Section \ref{Sec:Observations}, we describe the JWST observations and data reduction procedures, detailing the processing of NIRSpec IFU and MIRI MRS spectra and the calculation of observed [\ion{Fe}{II}] emission line intensities along the jet (Section \ref{Sec:Observations_observed_fe_intensities}). This is followed by a description of our excitation model for [\ion{Fe}{II}] in Section \ref{Sec:Fe2_model}. In Section \ref{Sec:4_Determine_Differences_in_Extinction}, we outline our method for determining relative extinction between gas emission lines, beginning with extinction differences in the simpler case of lines originating from the same upper energy level in the NIR (Section \ref{sec:4_NIR_extinction_Step1}). We then extend this analysis to the MIR (Section \ref{sec:4_NIR_vs_MIR_extinction}), using electron density constraints derived from NIR lines (Section \ref{sec:4_electron_density_NIR}) as a reference for comparing relative extinction. Finally, we consider cases where NIR and MIR lines trace different excitation conditions and examine how these differences influence the derived MIR extinction curves (Section \ref{sec:4_NIR_vs_MIR_extinction}). In Section \ref{Sec:Discussion}, we discuss the interpretations of our results in the context of existing extinction laws and dust grain evolution, and we conclude with a summary of our findings in Section \ref{Sec:Conclusions}.

\vspace{-9pt}
\section{Observations and Data Reduction}\label{Sec:Observations}

We present NIRSpec IFU and MIRI MRS observations of TMC1A. The NIRSpec data, originally introduced in \citet{harsono2023_tmc1a} (PID: 2104, PI: Harsono), focused on the [\ion{Fe}{II}] lines discussed in \citet{assani_tmc1a_asymmetry}. The MIRI MRS observations, presented here for the first time, were obtained on March 4, 2023, as part of the JWST Observations of Young ProtoStars (JOYS) Guaranteed Time Observations (GTO) program (PID: 1290, PI: E. van Dishoeck).

The NIRSpec IFU data, obtained on February 12, 2022, were reprocessed using JWST pipeline v1.11.4 \citep{bushouse_2023_1_11_4}, incorporating updated calibration files (JWST\_1123.PMAP) \citep{greenfield_calibration_reference_2016}. This version improved cosmic ray artifact handling, including snowball features, and enhanced flux calibration compared to previous reductions (v1.9.6) used in \citet{harsono2023_tmc1a}. Consistent with prior reductions, we skipped the Stage 3 outlier detection step to avoid excessive flagging of saturated pixels and strong emission lines \citep{Sturm_ERS_2023}. The final spectral cubes were manually inspected, and problematic spaxels were masked after continuum subtraction and applying signal-to-noise thresholds.

The MIRI MRS observations covered a wavelength range of 4.9–28.6 $\mu$m with a total integration time of 600 seconds across three sub-bands, each integrated with 36 groups in FASTR1 mode and two dithers optimized for extended emission. The data were processed using JWST pipeline v1.16.1 \citep{bushouse_2024_v_1_16_1}, with jwst1303.pmap as the calibration reference system. The Detector1Pipeline was applied to the uncal files with default settings to produce detector images, which were subsequently calibrated using the Spec2Pipeline. During this step, the fringe flat was subtracted from the data, residual fringe correction was applied, and a pixel-by-pixel background subtraction was performed using dedicated background detector images. A custom bad-pixel correction was applied using the VIP package \citep{christiaens_vip_python_2023_miri}. The final data cubes were created using the Spec3Pipeline with the drizzle algorithm \citep{Law_Drizzle_2023}, processing each band and channel independently. Due to a telescope mispointing, one of the two dither pointings in channel 1 did not cover the target, so only a single dither was used for that channel, while both dithers were used for all other channels. Background observations were taken immediately prior to the science exposures, using a single dither with 36 groups. The absolute flux calibration uncertainty for the MIRI-MRS observations is estimated to be 5.6\% $\pm$ 0.7\% \citep{argyriou_miri_performance_2023}.

To align the datasets, we determined the protostar’s position by fitting a 2D Gaussian function to the point spread function (PSF) in the NIRSpec and MIRI data. The PSF accounts for diffraction effects caused by JWST’s hexagonal mirror structure. For NIRSpec data, the centroid was calculated at wavelengths greater than 4$\mu$m using the G395H grating, while for MIRI data, centroids were derived independently for each spectral channel. These centroids were then used to spatially align the datasets, ensuring accurate correspondence of the protostar’s position across wavelengths. After alignment, the data were interpolated onto a common spatial grid using the 0.1\arcsec pixel scale of the NIRSpec IFU data. This resolution over-samples the native MIRI pixel scale (0.196–0.273\arcsec), but facilitates uniform region selection and comparison across wavelengths.

\vspace{-9pt}
\subsection{Observed Line Intensities}\label{Sec:Observations_observed_fe_intensities}

\begin{figure*}[ht]
    \centering
    \begin{subfigure}[b]{0.98\textwidth}
        \centering
        \includegraphics[width=\textwidth]{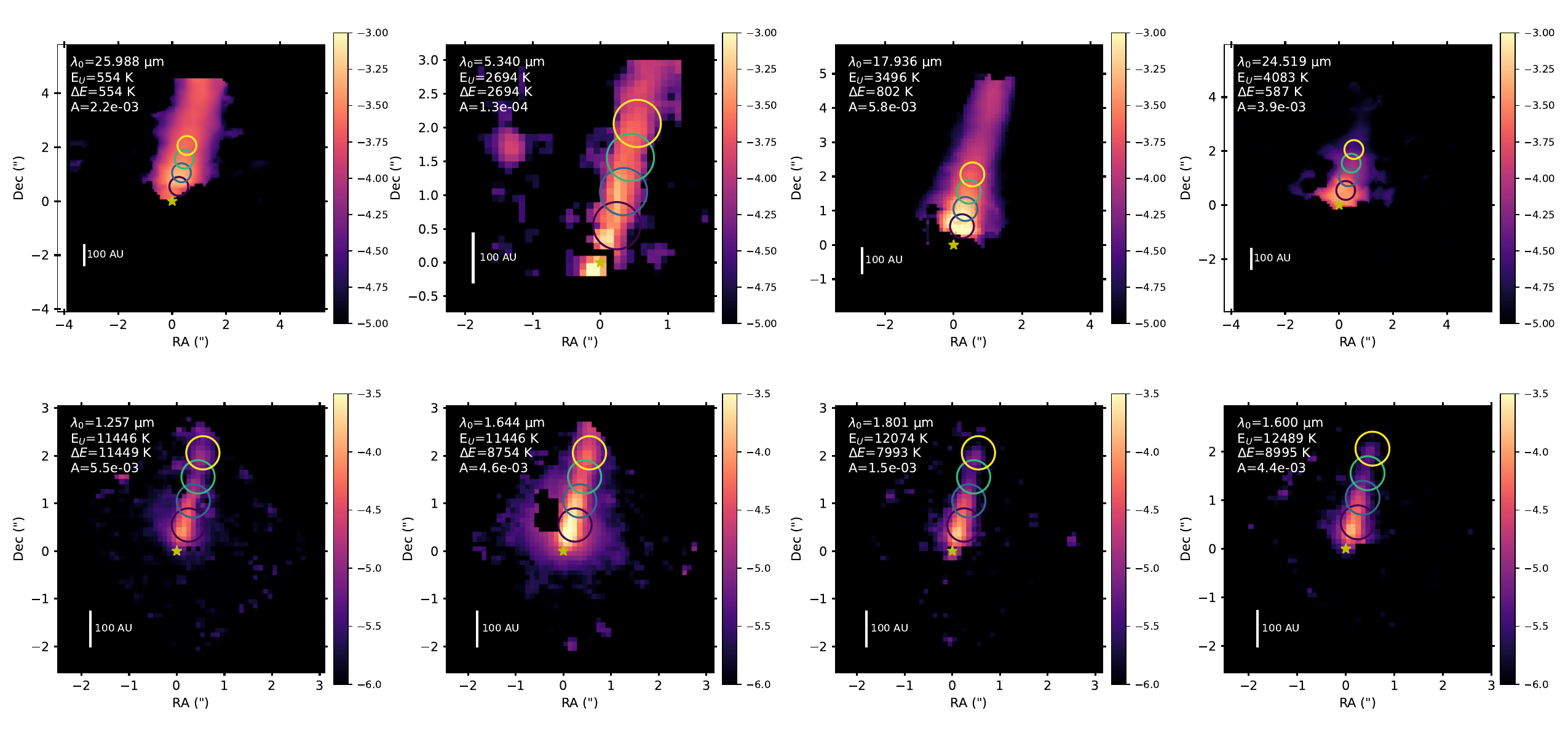}
    \end{subfigure}
    
    \vspace{-10pt} 
    
    \begin{subfigure}[b]{0.98\textwidth}
        \centering
        \includegraphics[width=\textwidth]{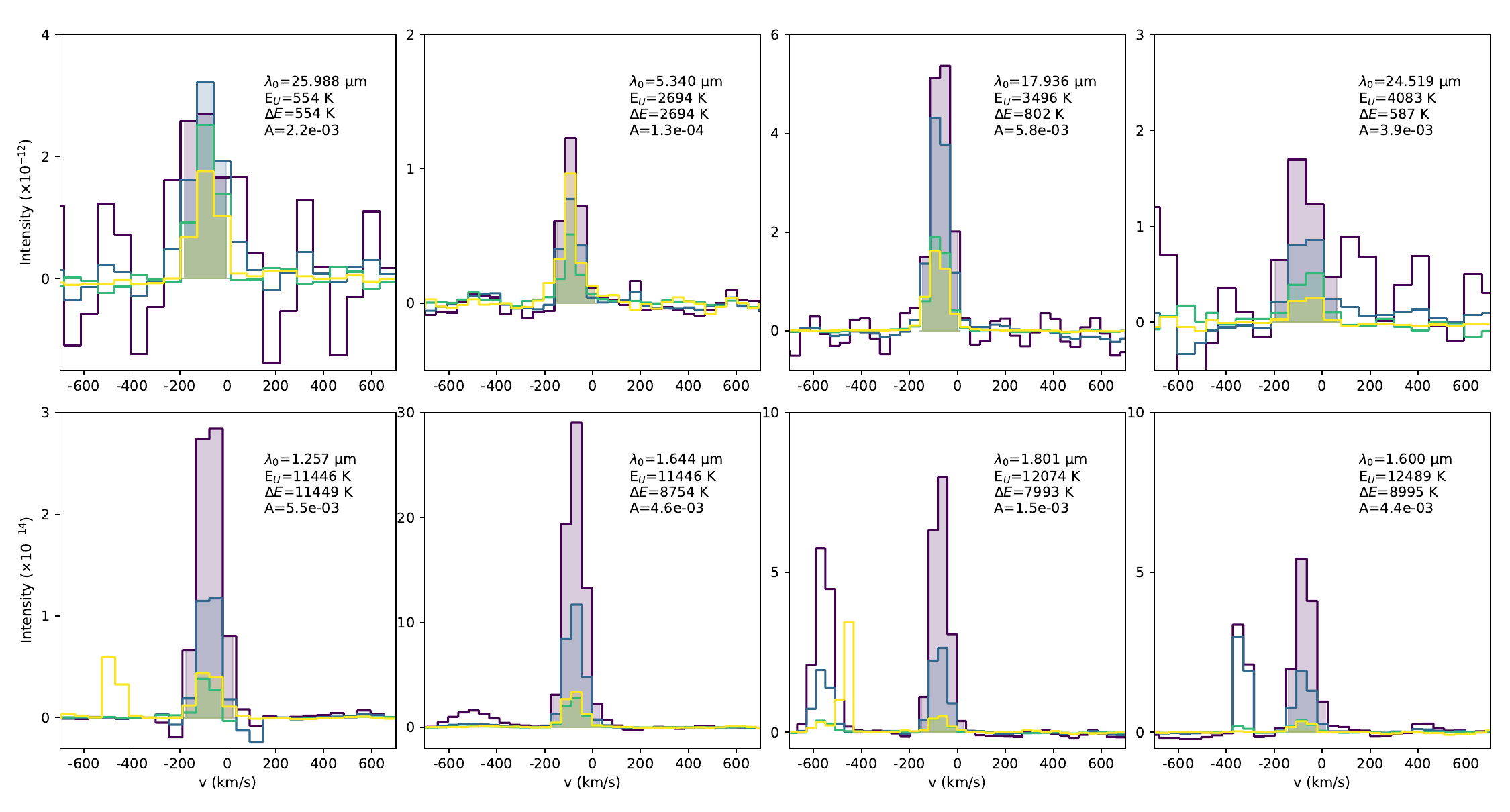}
    \end{subfigure}
    \vspace{-10pt}

    \caption{[\ion{Fe}{II}] emission lines used in this study and detected with NIRSpec and MIRI, arranged in order of increasing upper energy ($E_U/k$). \textbf{Top panels}: moment-0 maps (in units of erg s$^{-1}$ cm$^{-2}$ sr$^{-1}$), constructed by integrating the emission across the frequency range indicated by the shaded regions in the corresponding spectra below. Although integration is performed in frequency space, the x-axis is displayed in velocity (km s$^{-1}$) for clarity. A 3-sigma threshold is applied. The maps are overlaid with circular, color-coded apertures (0.7") used consistently throughout this study. \textbf{Bottom panels}: spectra (in units of erg s$^{-1}$ cm$^{-2}$ sr$^{-1}$ Hz$^{-1}$) extracted from each aperture, with shaded regions highlighting the velocity range used for integration, primarily focusing on blue-shifted emission. Each subplot includes the rest wavelength ($\lambda_0$), transition and upper energies ($\Delta E$, $E_U$), and Einstein-A value for reference.}
    \vspace{-10pt}

    \label{fig:iron_detections_subset}
\end{figure*}

This analysis focuses on [\ion{Fe}{II}] emission lines relevant to our excitation models (Section \ref{Sec:Fe2_model}), building on NIR detections reported in \citet{assani_tmc1a_asymmetry} and 4 new MIR [\ion{Fe}{II}] lines. 
In the NIR, we focus on the "a$^4$D" transitions which are detected across the jet and tend to be bright lines ($I_{\lambda}/I_{1.644}>10^{-2}$). We omit lines that are heavily blended---specifically, the 1.748 $\mu$m line (blended with H$_2$) and the 4.889 $\mu$m line (blended with CO v=1-0). The final dataset includes all "a$^4$D" transition detections along with the MIRI-detected lines 5.34, 17.99, 24.52 and 25.99 $\mu$m. A complete list of these lines, along with their atomic properties, including energy levels, Einstein A-coefficients, and critical densities, is provided in Table \ref{tab:spectroscopic_data}.

Observed line intensities were measured at four distinct locations along the blue-shifted axis of the TMC1A jet, focusing on the base of the jet (closer to the protostar) and the intensity peaks previously identified in \citet{assani_tmc1a_asymmetry}. These regions are shown as circular, color-coded apertures (diameter 0.7\arcsec) overlaid on the moment-0 maps in the top panels of Figure~\ref{fig:iron_detections_subset}. This aperture size exceeds MIRI’s native pixel scale (0.196–0.273\arcsec) and approximates instrument’s diffraction limit across the relevant wavelengths \citep[$\sim$0.2\arcsec at 5$\mu$m, $\sim$0.7\arcsec at 18$\mu$m, and $\sim$1\arcsec at 26$\mu$m;][]{Law_Drizzle_2023}, providing consistent spatial sampling across both NIRSpec and MIRI datasets while minimizing undersampling effects at longer wavelengths.

To extract line intensities, we summed over each aperture at every spectral channel to produce a 1D spectrum for each region (bottom panels of Figure~\ref{fig:iron_detections_subset}). Line intensities (in erg cm$^{-2}$ s$^{-1}$ sr$^{-1}$) were then computed by integrating the continuum-subtracted spectra in frequency space over $\sim$3–4 channels (corresponding to a velocity range of $\sim$150–200 km/s), tailored to encompass the full line while maintaining a consistent integration range across apertures for each line (shaded regions in the bottom panels of Figure~\ref{fig:iron_detections_subset}). The resulting intensities and line ratios (relative to the 1.644~$\mu$m line) are reported in Table~\ref{tab:integrated_intensities}.

\vspace{-9pt}
\section{The Iron Line Emission Model}\label{Sec:Fe2_model}

To model the intrinsic line intensities for [\ion{Fe}{II}], we use the CLOUDY spectral synthesis code \citep{ferland_cloudy_2017, chatzikos20232023}, which simulates physical and chemical processes in the ionized gas. We set up a single-zone calculation---assuming a uniform density and temperature---over a grid of gas densities and temperatures, considering radiative decay and collisional processes such as collisional excitation (where free electrons transfer kinetic energy to bound electrons, raising them to higher energy states) and collisional de-excitation (where collisions lower excited atoms to lower energy states). We disable photoionization and Ly$\alpha$ pumping in our model to remove contributions from radiative excitation, as their influence on [\ion{Fe}{II}] emission in jets remains uncertain. Since our technique for inferring extinction curves involves only ratios of [\ion{Fe}{II}] lines, which are controlled by the relative level populations of Fe$^+$, which, in turn, depend only on the density of electrons (the primary collider) and electron temperature $T_{\rm  e}$ in K, we follow \cite{verner1999numerical} and adopt solar abundances \citep{Grevesse_solar_abundances_2010} with a fixed, large Fe abundance (Fe/H = 100), a simple technique to leverage the complex machinery of the CLOUDY code to perform relatively straightforward excitation calculations that depend only on $n_{\rm e}$ and $T_{\rm e}$.  

For this study, we adopt the \cite{verner1999numerical} [\ion{Fe}{II}] spectroscopic dataset, which includes 371 energy levels and spans transitions from the infrared to the ultraviolet. CLOUDY defaults to the \cite{smyth2019towards} dataset, which extends to high temperatures for UV, and was found to best reproduce the observed UV and optical [\ion{Fe}{II}] spectra in the case of an AGN \citep{sarkar2021improved}. Differences in predicted line ratios between datasets arise from uncertainties in Einstein A-coefficients ($\sim$30–50\%) and collision strength calculations, which influence the final level populations. We find that \cite{verner1999numerical} reproduces near- and mid-IR [\ion{Fe}{II}] line ratios with electron density and temperature dependencies consistent with previous non-LTE models used for protostellar jets and Herbig-Haro (HH) objects \citep[e.g.,][]{giannini2015_HH1, koo_2016_shock_models}. Collisional strengths in the \cite{verner1999numerical} dataset for all quartet and sextet transitions are taken from \cite{Zhang_iron_collision_strengths_rates_1995} with transition rates taken from \cite{quinet_transition_probabilities_forbidden_FeII_1996}.
The model estimates a [\ion{Fe}{II}] line ratio of $I_{1.257}/I_{1.644} = 1.18$, which depends solely on the relative Einstein A-coefficients and transition frequencies, and remains constant across $n_{\rm e}$ and $T_{\rm e}$. This value is consistent with empirically derived intrinsic ratios of 1.11–1.20 \citep{giannini2015empirical}, and is commonly used to estimate extinction magnitudes \citep{assani_tmc1a_asymmetry, Erkaletal2021}. Therefore, we adopt the \cite{verner1999numerical} dataset, since a detailed comparison of infrared dataset performance is beyond the scope of this study.

We adopt a model grid with electron densities ($n_e \sim 10^3-10^6$ cm$^{-3}$) and temperatures (5,000–20,000 K), based on shock studies indicating that NIR and MIR [\ion{Fe}{II}] lines trace low-ionization gas with temperatures above 7,000 K and electron densities $n_e \sim 10^3-10^6$ cm$^{-3}$ \citep[e.g.,][]{Nisini_lecture_notes_2008, giannini2015_HH1, koo_2016_shock_models}. 

\vspace{-9pt}
\section{Differential Extinction in NIR and MIR}\label{Sec:4_Determine_Differences_in_Extinction}

We compare observed line ratios with intrinsic ratios from the excitation model to determine extinction differences between the NIR and MIR. Extinction modifies the observed ratio as:
\begin{equation}\label{eq:r_obs_to_r_model}
    r_{\rm obs}=  r_{\rm model}(n_e,T_e) \times 10^{-0.4  (A_{\lambda}-A_{\text{ref}})}
\end{equation}
where $r_{\rm obs}$ is the observed ratio between a NIR or MIR [\ion{Fe}{II}] line and the reference line, which in this study is the bright NIR line at 1.644 $\mu$m; $r_{\rm model}$ the intrinsic line ratio predicted by the excitation model as a function of $n_{\rm e}$ and $T_e$; and $A_{\lambda}$ and $A_{\rm ref}$ represent the extinction at the observed and reference wavelengths, respectively. Rearranging, the difference in extinction is given by:
 \begin{equation}\label{eq:delta_Alambda}
       \Delta A_{\lambda} \equiv A_{\lambda}-A_{\text{ref}} = -2.5 \log_{10}\left( \frac{r_{\rm obs}}{r_{\rm model}(n_e,T_e)} \right)
 \end{equation}
Since $\Delta A_{\lambda}$ depends only on $r_{obs}/r_{model}$, it directly quantifies the deviation between observed and intrinsic line emission ratios due to differential extinction relative to the reference wavelength. 

To compare the derived extinction differences with commonly used extinction curves, we normalize $\Delta A_{\lambda}$ by the extinction at the reference wavelength. The normalized extinction ratio is given by 
\begin{equation}\label{eq:beta}
    \beta_{\lambda} \equiv \frac{A_{\lambda}}{A_{ref}} = \frac{\Delta A_{\lambda}}{A_{ref}}+1
\end{equation} 
The reference extinction, $A_{1.644}$, is determined for each aperture along the jet using observed [\ion{Fe}{II}] integrated line ratios in the near-IR, where commonly used extinction curves show minimal deviation across 1-2 $\mu$m (see Sec \ref{Sec: introduction}). In this study, we use the bright 1.257/1.644 line pair, which arises from the a$^4$D J=7/2 fine-structure state, to calculate $A_{1.644}$ following equation 1 in \citet{assani_tmc1a_asymmetry}, using $\beta$ values ($\beta=A_{1.257}/A_{1.644}$) from established near-IR extinction curves \citep[e.g.,][]{cardelli1989relationship, Weingartner_and_Draine_01, pontoppidan_extinction_2024}. The resulting $A_{1.644}$ values for each aperture and extinction curve are listed in Appendix Table \ref{tab:extinction_magnitude}. Starting with the aperture closest to the protostar (purple, Fig. \ref{fig:iron_detections_subset}) and moving outward along the jet axis (yellow), we find $A_{1.644} = 4.11, 4.23, 4.00$, and 3.57 (corresponding to $A_V \sim 20$–17), with a standard deviation of $\sim$10\% between extinction curves and less than 1\% variation between R$_V$=3.1 and 5.5 within each curve.

To determine extinction differences between MIR lines and NIR lines (particularly the 1.644~$\mu$m line), we begin by establishing a robust reference extinction in the NIR. We first analyze extinction differences of NIR lines that share the same upper energy state as the 1.644 $\mu$m line (a$^4$D J=7/2), as their intrinsic ratios are independent of physical conditions. This allows us to derive $\Delta A_{\lambda}$ without uncertainty due to density or temperature effects. Next, we extend our analysis to NIR lines with different upper energy states, which provide a constraint on electron density ($n_{\rm e}$). This serves as a key reference for evaluating extinction in the mid-IR, where both $n_{\rm e}$ and $T_e$ influence the intrinsic line ratios. Finally, we compare NIR and MIR extinction properties and assess how potential differences in the physical conditions ($n_e, T_e$) among NIR and MIR emitting regions impact the interpretation of relative NIR/MIR extinction. 

\subsection{NIR Extinction Constraints Using Lines from the Same Upper Energy Level}\label{sec:4_NIR_extinction_Step1}

To begin, we validate our line-based extinction method using NIR [\ion{Fe}{II}] lines that share the same upper energy level as the 1.644 $\mu$m line (a$^4$D J=7/2). These line ratios are insensitive to variations in $n_{\rm e}$ and $T_{\rm e}$, making them ideal for isolating extinction effects. This provides a clean reference for computing $\Delta A_\lambda$ without uncertainty from physical conditions, before extending the analysis to lines with different upper states. 

\begin{figure}[ht]
    \centering
    \begin{subfigure}[b]{0.49\textwidth}
        \centering
        \includegraphics[width=\textwidth]{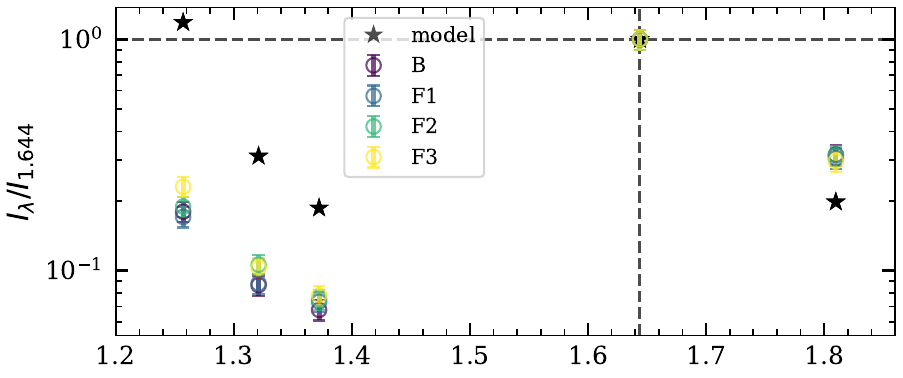}
    \end{subfigure}
    
    \begin{subfigure}[b]{0.46\textwidth}
        \centering
        \includegraphics[width=\textwidth, trim=0 0 15 0]{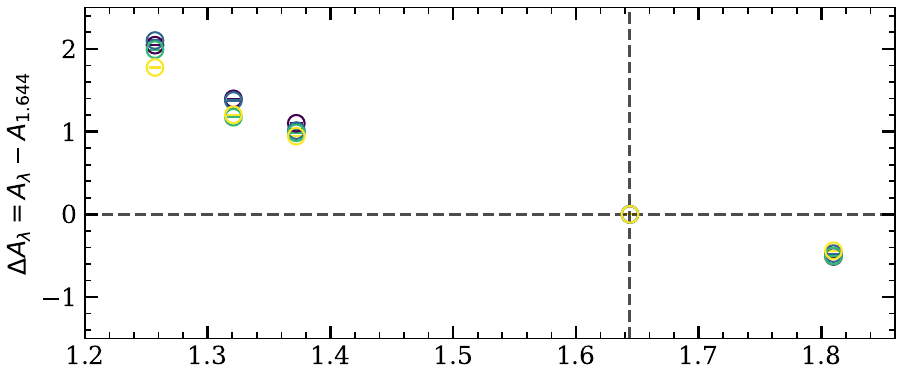}
    \end{subfigure}
    
    \begin{subfigure}[b]{0.49\textwidth}
        \centering
        \includegraphics[width=\textwidth]{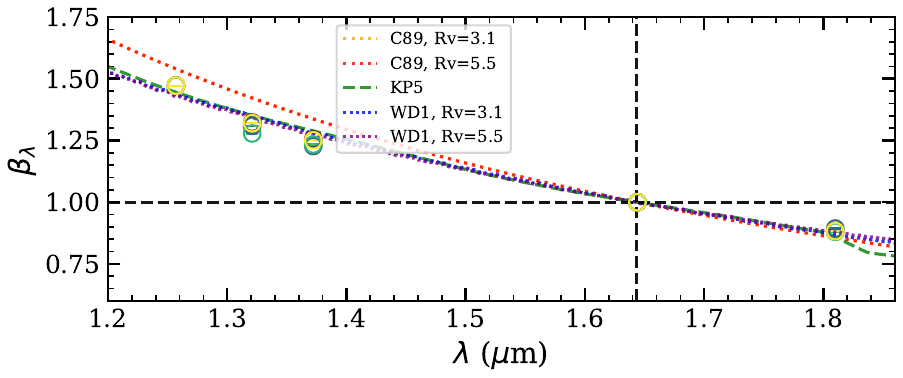}
    \end{subfigure}

    \caption{Near-infrared (NIR) extinction of a$^4$D J=7/2 lines relative to the 1.644 $\mu$m line (horizontal and vertical dashed lines.).  
\textbf{Top:} Observed (\texttt{'o'}) and modeled (\texttt{'*'}) line ratios for all four jet locations in this study.  
\textbf{Middle:} The difference in extinction relative to the 1.644 $\mu$m line.  
\textbf{Bottom:} Derived $\beta_{\lambda}$ values, normalized by the extinction at 1.644 $\mu$m, which is computed as an average over existing NIR extinction curves. This is compared with empirically determined NIR extinction curves---C89 \cite{cardelli1989relationship}, and modeled extinction curves---WD1 \cite{Weingartner_and_Draine_01}, KP5 \cite{pontoppidan_extinction_2024}.  
We include $R_V=3.1$ and $R_V=5.5$ for applicable curves to illustrate that NIR extinction in the 1.2–1.8 $\mu$m range is largely independent of $R_V$ and shows minimal differences between independent studies. Dashed lines mark the reference line ratio, extinction difference, and $\beta_{\lambda}$ value at 1.644 $\mu$m (horizontal) and the wavelength of 1.644 $\mu$m (vertical) in all three panels.}
    \label{fig:a4dj72_extinction}
\end{figure}

The top panel of Figure \ref{fig:a4dj72_extinction} presents the observed (\texttt{'o'}, not corrected for extinction) and the modeled (\texttt{'*'}) line ratios for the a$^4$D J=7/2 upper state across all four jet locations (see Figure \ref{fig:iron_detections_subset}). The middle panel shows the relative extinction differences ($\Delta A_{\lambda}$) computed using Equation \ref{eq:delta_Alambda}, referenced to the 1.644 $\mu$m line. Lines with $\lambda < 1.644$ $\mu$m (e.g., 1.257, 1.321, and 1.327 $\mu$m) exhibit positive extinction differences ($\Delta A > 0$), indicating greater extinction compared to the 1.644 $\mu$m line. Conversely, the 1.809 $\mu$m line has a negative extinction difference ($\Delta A < 0$), implying lower extinction. This confirms that the extinction curve in the 1–2 $\mu$m range has a negative slope, where shorter wavelengths suffer more extinction than longer wavelengths, consistent with previous near-IR extinction studies \citep[e.g.,][]{cardelli1989relationship, mathis_near_IR_extinction_1990, gordon_extinction_2023}. Additionally, the extinction differences exhibit spatial dependence across jet locations, with sightlines closer to the protostar ('B'/purple, 'F1'/blue) showing larger $\Delta A$ values at $\lambda < 1.644$ $\mu$m. This pattern aligns with the finding that total extinction magnitude decreases with distance from the protostar along sightlines toward the TMC1A jet, as demonstrated in \cite{assani_tmc1a_asymmetry}. Consequently, normalizing by the total extinction at each jet location should account for these differences, revealing a more intrinsic extinction trend across locations.

The negative slope observed in our derived $\Delta A_{\lambda}$ values (middle panel) and the close agreement between our derived $\beta_{\lambda}$ values and independent extinction curves and models (bottom panel) confirm the consistency of the extinction curves in the 1.2–1.8 $\mu$m range. This consistency reinforces the reliability of using a gas line in this range---such as 1.644 $\mu$m---as a reference for measuring relative extinction differences. Moreover, having established a well-constrained extinction correction in the NIR, we can now use these NIR lines to probe the physical conditions of the gas emitting these lines, specifically placing constraints on the electron density ($n_{\rm e}$).

\vspace{-10pt}
\subsection{Electron Density Constraints from NIR Lines}\label{sec:4_electron_density_NIR}

\begin{figure}[!htbp]
    \centering
    \begin{subfigure}[b]{0.48\textwidth}
        \centering
        \includegraphics[width=\textwidth]{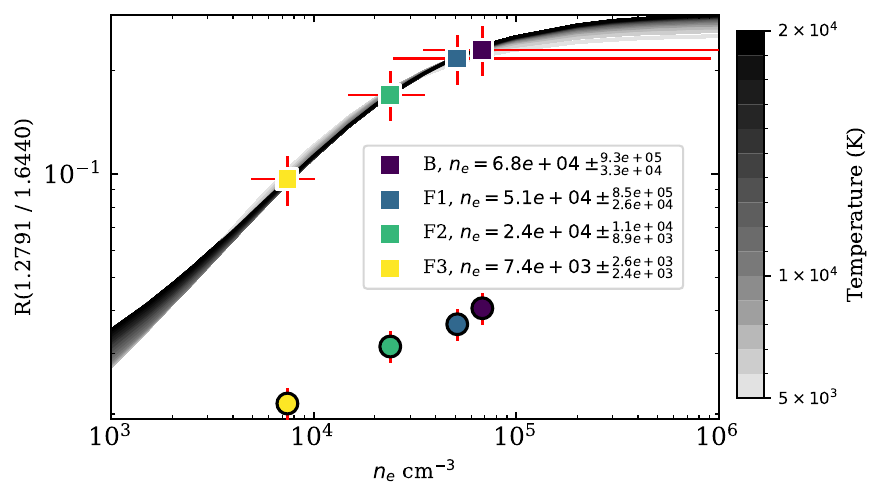}
    \end{subfigure}
    
    \begin{subfigure}[b]{0.48\textwidth}
        \centering
        \includegraphics[width=\textwidth]{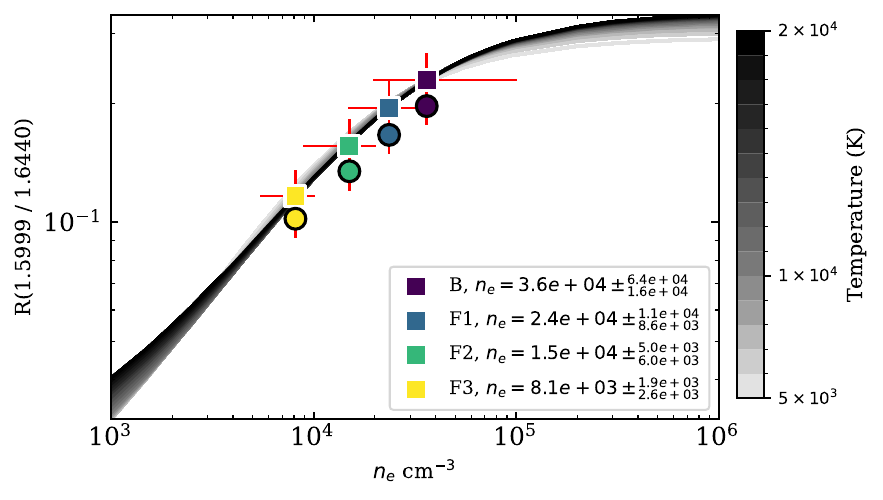}
    \end{subfigure}
    
    \begin{subfigure}[b]{0.48\textwidth}
        \centering
        \includegraphics[width=\textwidth]{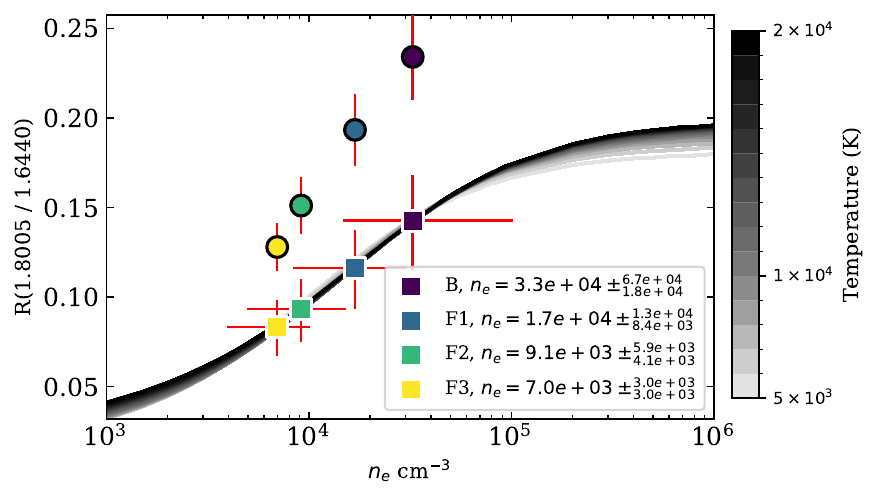}
    \end{subfigure}

    \caption{Examples of density-sensitive NIR line ratios, where the CLOUDY model predictions are shown as solid lines shaded by temperature. Observed line ratios with uncertainties are represented by circles, while extinction-corrected ratios are shown as squares color-coded by jet location (see Fig. \ref{fig:iron_detections_subset}). The extinction-corrected ratio, accounting for uncertainties, determines the electron density from the model, aligning with the x-axis position of both observed and corrected ratios. \textbf{Top}: A shorter-wavelength line is more affected by extinction, leading to an increased ratio after correction. \textbf{Middle}: A line ratio with similar rest wavelengths, minimally affected by extinction, where the observed ratio closely matches the expected electron density. \textbf{Bottom}: A longer-wavelength line less affected by extinction, causing the ratio to decrease after correction. Following extinction correction using established NIR curves (Section \ref{sec:4_NIR_extinction_Step1}), all corrected ratios fall within the model-predicted range, showing consistent electron density trends across jet locations and line pairings. This electron density serves as a starting point for comparing relative extinction in the NIR/MIR (Section \ref{sec:4_NIR_vs_MIR_extinction}).}
    \label{fig:ne_a4D_subset}
\end{figure}

We now extend our analysis to all observed a$^4$D transitions in the near-IR to constrain the electron density ($n_{\rm e}$) at each jet location (see Figure \ref{fig:iron_detections_subset}). The line ratios of a$^4$D transitions, which originate from states with similar but slightly different upper-level energies ($E_{\rm upper} = 11,446 - 12,729$ K) and distinct critical densities ($n_{\rm crit} \approx 4$–$7 \times 10^4$ cm$^{-3}$ at $T = 10,000$ K), are relatively insensitive to temperature but remain sensitive to electron densities in the $10^{3\text{–}5}$ cm$^{-3}$ range. This makes them well suited for quantifying $n_{\rm e}$ in the near-IR, where extinction curves are well established.

The derived electron density is not only important in its own right (as a key physical quantity of the NIR line emitting gas in the jet), but it also serves as a starting point for discussing the electron density in the MIR line emitting region, which is unknown without a firm knowledge of the extinction law for the protostellar envelope where the jet is embedded and may or may not be the same as that in the NIR line emitting gas. This distinction is particularly relevant as the MIR lines arise from lower excitation levels and may trace gas under different physical conditions (see Section \ref{sec:4_NIR_vs_MIR_extinction} below). 

Figure \ref{fig:ne_a4D_subset} presents three examples NIR [\ion{Fe}{II}] a$^4$D line ratios used to determine electron density. The predicted line ratios from the [\ion{Fe}{II}] model are plotted as a function of electron density across temperatures from 5,000 to 20,000 K. Observed ratios (circles) and extinction-corrected ratios (squares) are shown with uncertainties. We adopt a constant $\sim$15\% uncertainty in extinction for simplicity, consistent with the typical deviations between near-IR extinction curves (see Table \ref{tab:extinction_magnitude}) and uncertainties in the Einstein A-values \citep{giannini2015empirical}, which result in larger uncertainties for the corrected ratios than for the observed ones. Electron density uncertainties primarily arise from (1) uncertainty in the intrinsic line ratios, broadening the possible $n_{\rm e}$ range by approximately a factor of two, and (2) temperature-dependent differences in electron density for a given ratio, which are typically small (<5\%). 

Near the protostar, electron density estimates approach the critical density ($n_{\rm crit}$) of these transitions. Since line ratios become insensitive to the density at $n_e \gtrsim n_{\rm crit}$ due to thermalization effects (as the gas approaches LTE conditions), constraining electron densities in such regions becomes increasingly uncertain. As a result, the density estimates in the bright region ('B'/purple) should be interpreted as lower limits. Further from the protostar, intensity ratios relative to the 1.644 $\mu$m line are lower, leading to consistently lower and well-constrained electron density estimates.

Across 13 NIR a$^4$D line ratios, electron density estimates remain internally consistent, with differences limited to a factor of $\sim$2 at each jet location (see Table \ref{table:electron_densities_a4D}). The results show a clear trend of decreasing electron density along the jet axis: $\bar{n}_e=(3.6\pm 1.0)\times10^4$ cm$^{-3}$ (knot ``B"/purple), $(2.4\pm 0.6)\times10^4$ cm$^{-3}$ (``F1"/blue), $(1.5\pm 0.5)\times10^4$ cm$^{-3}$ (``F2"/green), and $(8.0\pm 3.0)\times10^3$ cm$^{-3}$ (``F3"/yellow). While uncertainties in adjacent regions overlap in some cases, the overall trend of decreasing electron density with distance from the protostar remains robust. The electron density drops between the bright region (B) and the third peak (F3) exceeds observational uncertainties in most ratios, further reinforcing this trend.

Now that the electron density has been determined using NIR [\ion{Fe}{II}] gas lines at each jet location, we can use these values as reference points to assess relative extinction between the NIR and MIR along each line of sight toward the TMC1A jet.

\vspace{-9pt}
\subsection{Relative Extinction between NIR and MIR}\label{sec:4_NIR_vs_MIR_extinction}

To determine the relative extinction between the NIR and MIR, we compare observed [\ion{Fe}{II}] line ratios of MIR transitions (a$^6$D, a$^4$F at 5.3, 17.0, 24.5, and 26 $\mu$m) relative to the 1.644 $\mu$m line with intrinsic model ratios. We first assume that the NIR and MIR lines along the same sightlines originate from the same physical region, implying identical excitation conditions governed by the same $n_{\rm e}$ and $T$. Next, we explore the case where MIR lines along the same sightlines trace gas with conditions different from NIR lines, with different values of $n_{\rm e}$ and/or $T$. 

The electron density derived from NIR lines (Sect. \ref{sec:4_electron_density_NIR}) serves as the reference for these comparisons at each jet location. Uncertainties in electron density and temperature contribute to the overall uncertainty range. We adopt a standard reference temperature of 10,000 K, consistent with shock models, where NIR and MIR lines become emissive at temperatures $\gtrsim$ 7,000 K \citep{Nisini_lecture_notes_2008, giannini2015_HH1, koo_2016_shock_models}. To account for uncertainties, we use a lower bound of 5,000 K.

\vspace{-9pt}
\subsubsection{Case 1: Same Physical Conditions}\label{sec:4_NIR_vs_MIR_extinction_case1}

In the scenario where the NIR and MIR [\ion{Fe}{II}] emission lines along the same sightlines originate from the same physical region, we first calculate the modeled line ratios assuming the same electron density and temperature for both MIR and NIR lines, including the reference line at 1.644~$\mu$m. 
\begin{figure*}[t]
    \centering
    \begin{subfigure}[b]{0.48\textwidth}
        \centering
        \includegraphics[width= \linewidth, trim=3mm 0mm 2mm 0mm]{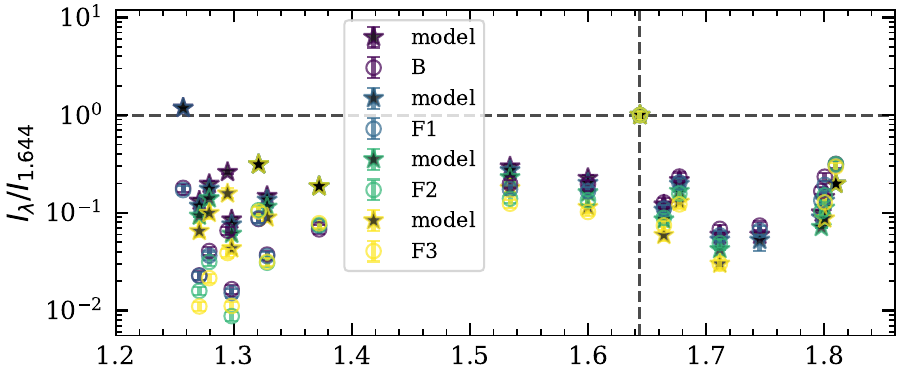}
    \end{subfigure}
    \hfill
    \begin{subfigure}[b]{0.48\textwidth}
        \centering
        \includegraphics[width=\linewidth, trim=1mm 0mm 5mm 0mm]{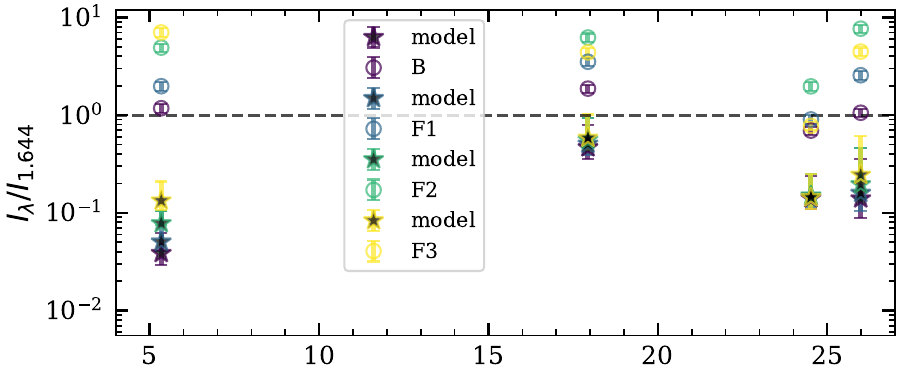}
    \end{subfigure}
    \hfill
    \vspace{0.3cm}
    \begin{subfigure}[b]{0.43\textwidth}
        \centering
        \includegraphics[width=\linewidth,trim=0mm 0mm 19mm 0mm]{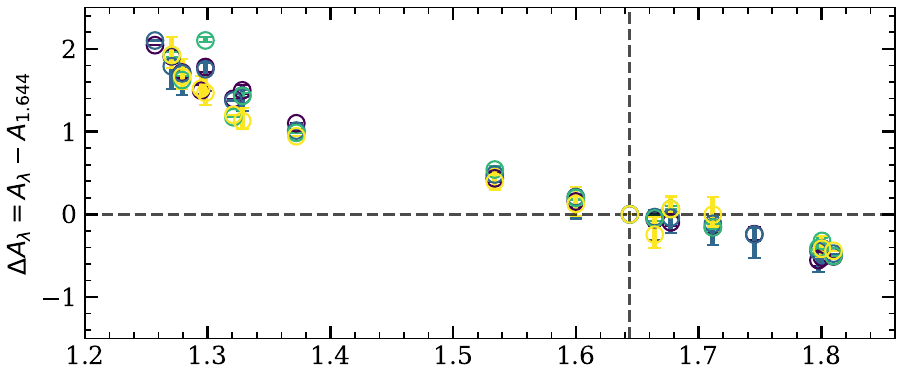}
    \end{subfigure}
    \hfill
    \begin{subfigure}[b]{0.44\textwidth}
        \centering
        \includegraphics[width=\linewidth, trim=13mm 0mm 0mm 0mm]{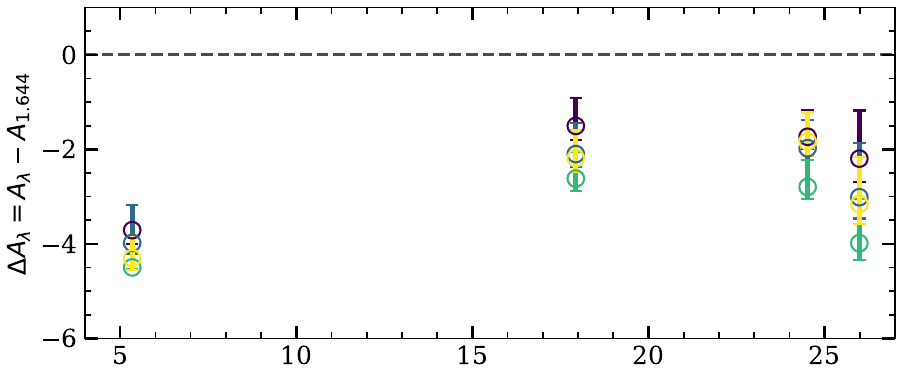}
    \end{subfigure}
    \vspace{0.3cm}
    \begin{subfigure}[b]{0.49\textwidth}
        \centering
        \includegraphics[width=\linewidth]{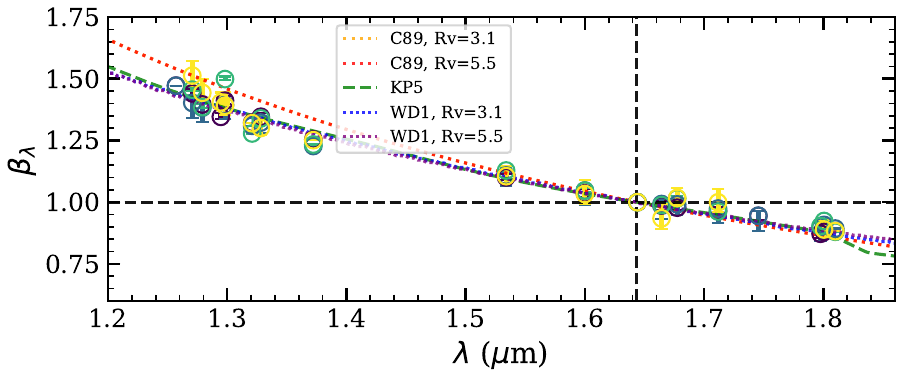}
    \end{subfigure}
    \hfill
    \begin{subfigure}[b]{0.49\textwidth}
        \centering
        \includegraphics[width=\linewidth]{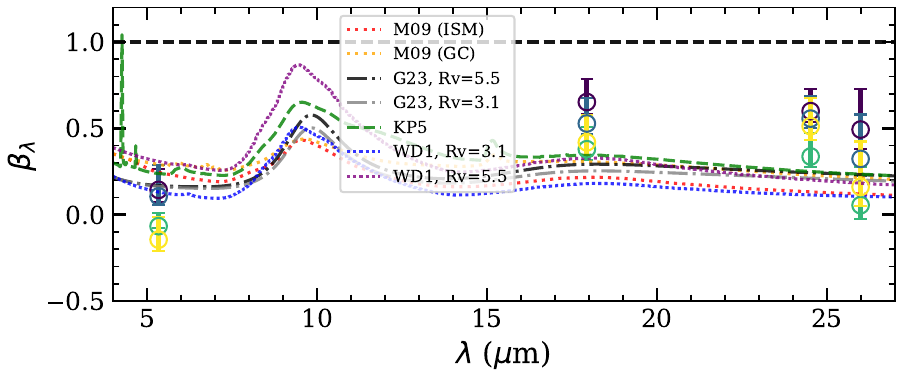}
    \end{subfigure}
    

    \caption{\textit{NIR and MIR extinction toward the TMC1A jet}. Same as Figure 2, but extended to include additional a$^4$D lines in the NIR (left) used for $n_{\rm e}$ derivation, along with MIR lines (right). Model scatter points correspond to the derived electron density at 10,000 K (see Sect. \ref{sec:4_electron_density_NIR}), with uncertainties reflecting the uncertainty range in the determined $n_{\rm e}$ value and temperatures from 5,000–10,000 K. Commonly used NIR/MIR extinction curves are overplotted for comparison (see text).}
    \label{fig:MIR_derived_extinction_same_cond}
\end{figure*}

Figure \ref{fig:MIR_derived_extinction_same_cond} shows the derived extinction for each line of sight toward the TMC1A jet. The left panels show NIR-derived extinction in the 1.2–1.85 $\mu$m range, including additional a$^4$D lines not used to derive the electron density, which follows the trend observed in Sect. \ref{sec:4_NIR_extinction_Step1}. The right panels show extinction derived from the four MIR lines observed with MIRI. 

The observed MIR line fluxes are comparable to or higher than that of the 1.644~$\mu$m line (with ratios $r_{\mathrm{obs}}\sim$1–10), which is very different from the intrinsic line ratios modeled, which are below unity (with $r_{\mathrm{model}} < 1$). The difference is caused by the fact that the reference NIR line at 1.644~$\mu$m experiences stronger extinction than the MIR lines, which is reflected by the negative $\Delta A_{\lambda}$ shown in the middle right panel of Fig.~\ref{fig:MIR_derived_extinction_same_cond}.

Extinction at each jet location follows a pattern where bright regions (B, F1) exhibit higher MIR extinction than regions farther from the protostar (F2, F3), consistent with increased column density closer in. Our inferred extinction at 5.34 $\mu$m is low ($\beta_{\lambda} \sim 0$), likely setting a lower limit since it is comparable to or less than the least extinct curve at that wavelength (WD1, $R_V=3.1$). In contrast, extinction at 17.93, 24.52, and 25.99 $\mu$m is usually higher than the comparison curves, indicating greater attenuation at these wavelengths compared to that expected from the comparison curves. 

Since $r_{\rm obs} > r_{\rm model}$ for the MIR lines, we find $\Delta A_{\lambda} < 0$, and any additional suppression of the observed MIR intensity due to line blending would only make $\Delta A_{\lambda}$ less negative, thereby increasing $\beta_{\lambda}$. This could be the case for the [\ion{Fe}{II}] 17.936 $\mu$m line, which is spectrally close to the H$_2$ $v=1$–1 S(1) line at 17.94 $\mu$m. However, as a first-order approximation, we built an H$_2$ excitation model and interpolated the column density of H$_2$ molecules in the upper state ($v_u = 1$, $J_u = 3$) based on the energy of the corresponding upper state of H$_2$ 1-1 S(1). We then estimated the H$_2$ 1-1 S(1) line flux using the Einstein coefficient of spontaneous emission, assuming an extinction law (KP5; \cite{pontoppidan_extinction_2024}) and corresponding extinction value. This allowed us to calculate the fraction of the observed 17.936 $\mu$m line that originates from H$_2$ and [\ion{Fe}{II}]. We find the H$_2$ contribution is $\ll$1\%, confirming that [\ion{Fe}{II}] dominates at this wavelength.

\subsubsection{Case 2: Diverging Physical Conditions between NIR and MIR lines}\label{sec:4_NIR_vs_MIR_extinction_case2}

The observed [\ion{Fe}{II}] lines may originate from gas with different excitation conditions. This posits that the emitting regions are stratified, where higher-excitation lines (e.g., those in the NIR) predominantly trace hotter and/or denser gas, while lower-excitation lines (e.g., those in the MIR) arise from cooler and/or less dense regions along the same line of sight (within the effective telescope beam). 

To explore this possibility, we consider a simplified scenario in which the NIR ($\lambda<5~\mu$m) and MIR ($\lambda>5~\mu$m) lines trace two distinct temperature and density components, with
\begin{equation} r_{\text{model}} = \frac{I_{\lambda_{\text{MIR}}}(f_{n_e} n_{e}, f_T T_{e})}{I_{1.644}(n_{e}, T_{e})} \end{equation}
where $f_{n_e}$ and $f_T$ represent the fractional electron density and temperature of the MIR-emitting gas relative to those inferred from the NIR a$^4$D line ratios (see Sect. \ref{sec:4_electron_density_NIR}).

\begin{figure}[htbp]
    \centering
    \begin{subfigure}[b]{0.47\textwidth}
        \centering
        \caption{$f_T=0.5, f_{n_e}=1.0$}
        \includegraphics[width=\linewidth]{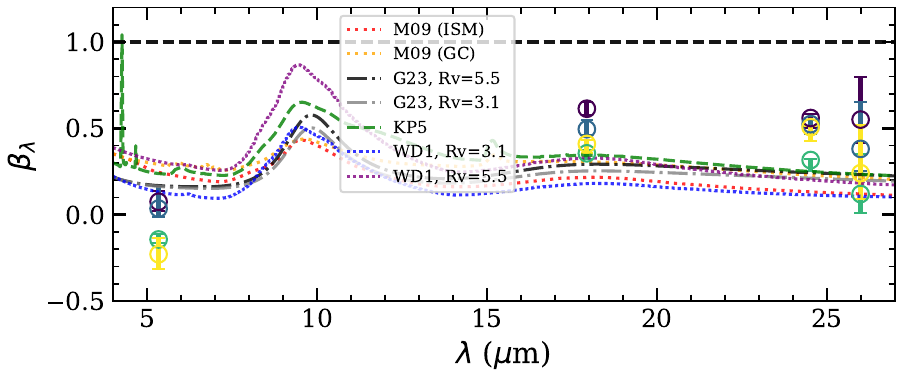}
    \end{subfigure}
    \begin{subfigure}[b]{0.47\textwidth}
        \centering
        \caption{$f_T=1.0, f_{n_e}=0.5$}
        \includegraphics[width=\linewidth]{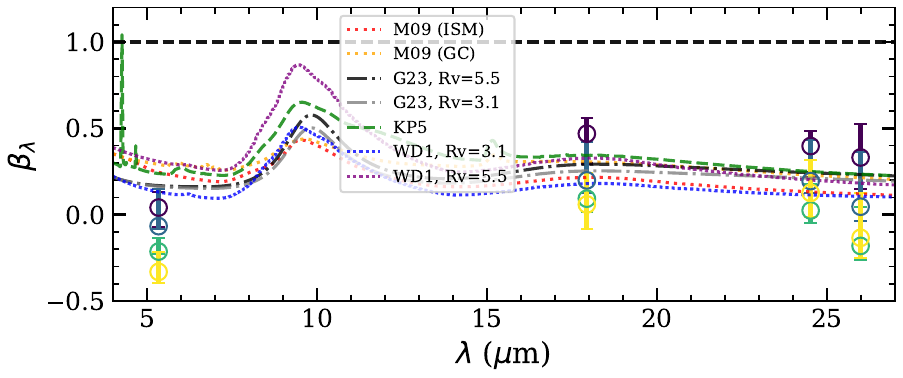}
    \end{subfigure}
    \begin{subfigure}[b]{0.47\textwidth}
        \centering
        \caption{$f_T=0.5, f_{n_e}=0.5$}

        \includegraphics[width=\linewidth]{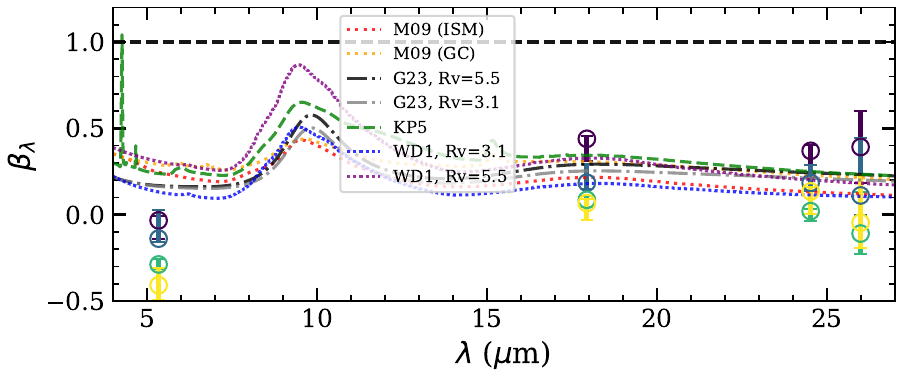}
    
    \end{subfigure}
    \caption{Derived MIR extinction from [\ion{Fe}{II}] lines ($\lambda >5$ $\mu$m) tracing different excitation conditions compared to NIR [\ion{Fe}{II}] lines. The plots show MIR extinction, where MIR [\ion{Fe}{II}] lines trace different excitation conditions ($n_e, T_e$) than NIR [\ion{Fe}{II}] lines. \textbf{Top} $\beta_{\lambda}$ values assuming MIR line intensities correspond to gas at half the NIR-traced temperature ($f_T=0.5$) with the same electron density ($f_{n_e}=1$). \textbf{Middle} Same temperature as NIR-traced gas ($f_T=1.0$), but assuming MIR lines trace half the electron density ($f_{n_e}=0.5$). \textbf{Bottom} Both temperature and electron density are halved ($f_T=0.5$, $f_{n_e}=0.5$).}
    \label{fig:MIR_extinction_diff_cond}
\end{figure}

For temperatures above 5,000 K, predicted line ratios converge at a given electron density, resulting in minimal temperature dependence on the derived $\beta_{\lambda}$ values. This is seen in the top subplot of Fig.~\ref{fig:MIR_extinction_diff_cond}, where the MIR lines are assumed to originate from gas at half the temperature of the NIR lines ($f_T = 0.5$, $T_{\mathrm{MIR}} = 5,000$ K, $T_{\mathrm{NIR}} = 10,000$ K). The derived MIR $\beta_{\lambda}$ values are almost indistinguishable from the reference case of $f_T = 1$ and $f_{n_e}=1$ shown in the lower-right panel of Fig.~\ref{fig:MIR_derived_extinction_same_cond}.

In contrast, electron density has a more pronounced effect on $\beta_{\lambda}$. The middle subplot ($f_T = 1.0$, $f_{n_e} = 0.5$) shows that lowering the electron density systematically reduces $\beta_{\lambda}$, consistent with an increased MIR/NIR line ratio that decreases the relative difference between observed and modeled values. The bottom subplot ($f_T = 0.5$, $f_{n_e} = 0.5$) shows a similar $\beta_{\lambda}$ profile to the middle case, reinforcing that the electron density difference has a stronger influence on the derived extinction than temperature differences.

\section{Discussion}\label{Sec:Discussion}
\begin{figure}
    \centering
    \includegraphics[width=1.0\linewidth,trim=0cm 0mm 0mm 0mm]{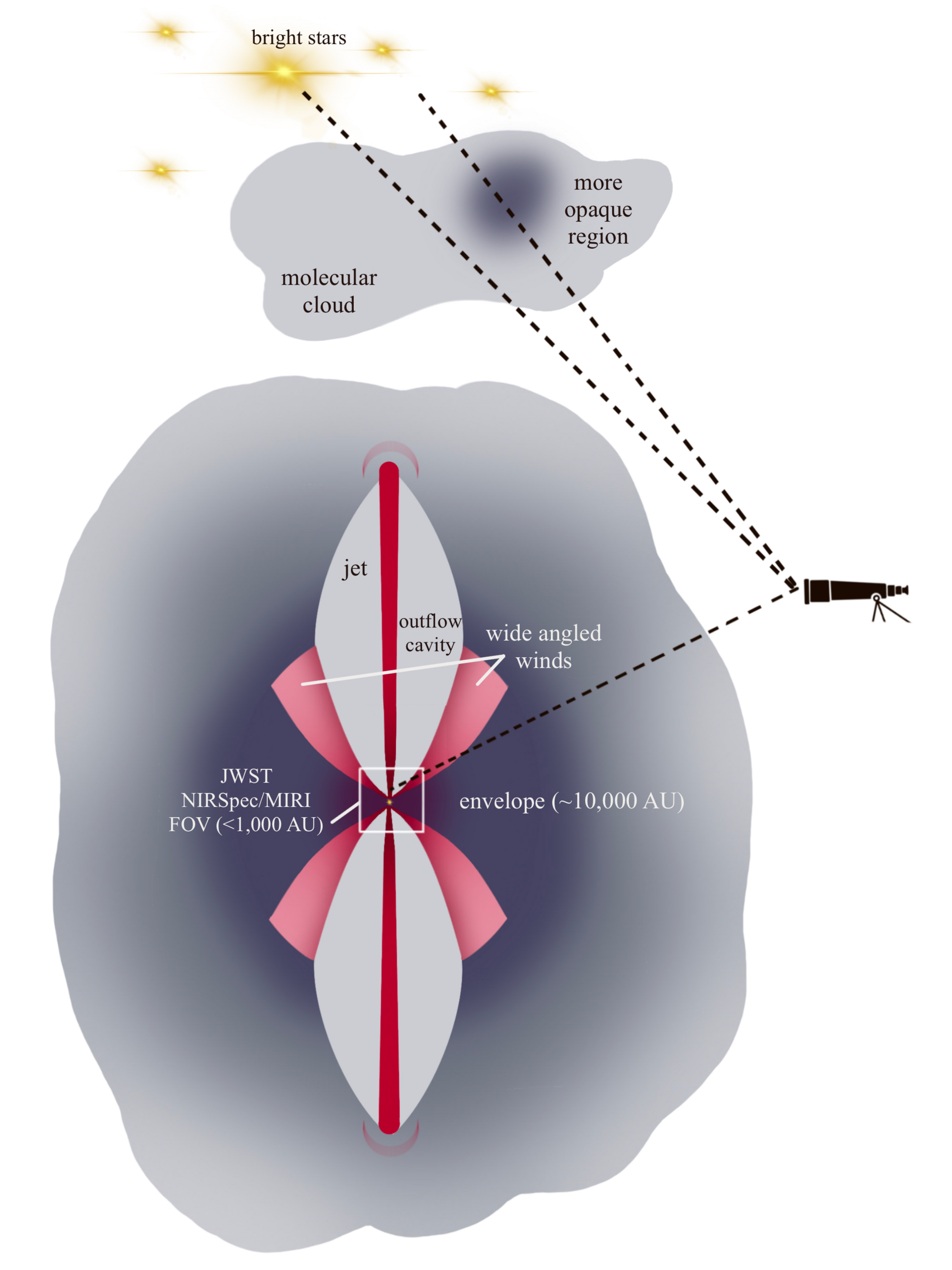}
    \caption{Schematic illustrating extinction geometry in two contexts: (top) background star extinction studies, and (bottom) protostellar envelopes. The bottom panel (adapted from Fig. 5 of \cite{ewine_wish_survey1}) shows a forming star embedded in a large infalling envelope, with bipolar jets carving outflow cavities. The JWST NIRSpec/MIRI field of view probes $<$1000 AU scales, with dashed lines tracing the line-of-sight path through the dense envelope, where most extinction occurs. In contrast, the top panel illustrates background stars viewed through diffuse and dense regions of a molecular cloud \citep[e.g.][]{mcclure_extinction_law_2009}, probing more pristine dust that has not yet experienced collapse or been processed by star formation. While the top probes molecular cloud dust, the embedded jet sightlines (bottom) trace extinction through evolved, denser material. Higher MIR extinction near the protostar---if MIR and NIR lines trace similar gas---may indicate a change in the grain size distribution potentially due to further grain growth as the cloud condenses and collapses to form the central protostar.}
    \label{fig:extinction_schematic}
\end{figure}

We developed a new technique for determining the MIR extinction in the protostellar envelope using JWST IFU observations of several [\ion{Fe}{II}] NIR and MIR emission lines, leveraging the well-determined NIR extinction curve and excitation model for line ratios. It is applied to the atomic jet toward Class I protostar TMC1A which is embedded in a large scale infalling envelope. This technique provides a spatially resolved constraint on extinction along different sightlines to the jet, offering an independent measurement distinct from background star methods applied to diffuse and dense clouds away from protostars \citep[e.g.,][]{mcclure_extinction_law_2009}; it complements opacity measurements of volatile ice features in protostellar envelopes along similar sightlines. \citep[e.g.,][]{Tyagi_icemapping_jwstIPA_hops370_2024}. Figure \ref{fig:extinction_schematic} illustrates this distinction, with the top panel showing extinction measured from background stars and the bottom panel representing embedded jet sightlines that pass through a large and dense, infalling envelope.  

If the MIR lines originate from gas in the embedded jet with physical conditions (electron density $n_{\rm e}$ and temperature $T$) similar to those of the NIR line emitting region along the same sightlines, we find systematically higher MIR extinction through the protostellar envelope compared to the the empirical high-extinction regime derived by \cite{mcclure_extinction_law_2009}. It also exceeds the R$_V$=5.5 case from \cite{Weingartner_and_Draine_01} and the KP5 model from \cite{pontoppidan_extinction_2024} (see the lower-right panel of Fig.~\ref{fig:MIR_derived_extinction_same_cond}).  These models describe dust populations with larger maximum grain sizes than those in the diffuse ISM. The alignment of \cite{mcclure_extinction_law_2009} with the R$_V$=5.5 case has been interpreted as evidence for dust grain growth in denser cloud regions. If the MIR and NIR emission lines trace the same excitation conditions, the higher inferred extinction in the MIR toward the embedded jet suggests differences in the grain size distribution such as the presence of even larger dust grains along these sightlines.
This result adds weight to the conclusion of \citet{Valdivia_polarization_grain_growth_2019} and \cite{Le_Gouellec_dust_polarization_2019} that dust grains must have grown significantly in protostellar envelopes---since relatively long-wavelength local IR photons can only spin them up efficiently through radiative torques if the grains are sufficiently large, enabling magnetic alignment and the observed levels of dust polarization.   

If the MIR emission lines trace lower electron density and temperature than the NIR lines, then the modeled intrinsic MIR lines would be weaker, yielding a lower inferred MIR extinction, with the resulting $\beta_{\lambda}$ values falling within the range predicted by background star studies (see Sec \ref{sec:4_NIR_vs_MIR_extinction_case2}).  In this case, the extinction curve does not indicate significant changes in grain size distributions such as dust growth beyond what is inferred along high-extinction sightlines of dense clouds. This result is still highly significant because it implies that, even along the same sightlines to the embedded jet, the MIR lines must originate from gas with properties different from those of the NIR line emitting gas. This could happen if, e.g., the MIR and NIR lines trace different regions behind spatially unresolved shocks in the telescope beam. Differences in excitation conditions between NIR and MIR lines have been observed in the case of H$_2$ outflows \citep[e.g.,][]{Vleugels_H2jet_ClassI_2025}, but to our knowledge, this has not yet been demonstrated for atomic lines from the same ionization state. Our result highlights the important point that it would be difficult to accurately infer the physical properties of the MIR line-emitting gas in the embedded jet unless the MIR extinction curve in the foreground protostellar envelope is well characterized independently.  

The method described in this paper, leveraging wide spectral coverage and integral field unit (IFU) observations, could be extended to other embedded astrophysical environments. Atomic and molecular tracers, such as forbidden lines, hydrogen recombination lines, H$_2$ with their wide spectral coverage across infrared and radio wavelengths, offer promising opportunities for applying this technique. If these tracers originate from the same spatial region and their line intensities or ratios can be reliably modeled, extinction curves could be determined. Additionally, applying this method to species tracing spatially distinct regions (e.g., spatially nested outflows) and comparing them self-consistently within the same source could provide deeper insights into the dust properties across these regions, offering a unique perspective on the role of dynamic processes in star formation (particularly outflows) in shaping dust. Expanding this approach to more sources at different stages of star formation and along diverse lines of sight would enhance our understanding of dust evolution within protostellar environments and throughout star formation. 

\section{Conclusions}\label{Sec:Conclusions}

This study investigates the extinction properties and evidence of dust processing in the protostellar envelope surrounding the embedded [\ion{Fe}{II}] jet of the Class I protostar TMC1A using JWST NIRSpec and MIRI observations and a novel technique. By combining high-sensitivity, multi-wavelength emission line data with excitation models, we derived localized extinction curves along sightlines to the embedded jet and compared them to extinction laws derived from dark cloud populations. These efforts aimed to understand how star formation processes influence the surrounding dust populations, contributing to differences in line-of-sight extinction. We summarize our findings as follows:

\begin{enumerate}
    \item We have developed a new method for characterizing the NIR-MIR extinction of the material in the protostellar envelope along the lines of sight to an embedded [FeII] jet based on JWST NIRspec and MIRI observations of line emissions and excitation model predictions of intrinsic line ratios. 
    \item The modeled intrinsic line ratios depend on electron density and temperature. The electron density can be reasonably well constrained by NIR [\ion{Fe}{II}] lines (for which extinction curves are already well characterized), with values ranging from approximately 5 × 10$^3$ to 5 × 10$^4$ cm$^{-3}$, decreasing farther from the protostar, consistent with prior studies of protostellar outflows. 
    
     \item If the MIR lines originate from gas in the embedded jet with electron density $n_{\rm e}$ and temperature $T$ similar to those of the NIR line emitting region along the same sightlines, we find a systematically higher MIR extinction through the protostellar envelope than those derived by \cite{mcclure_extinction_law_2009} for dark clouds, the R$_V$=5.5 case modeled by  \cite{Weingartner_and_Draine_01}, and the KP5 model from \cite{pontoppidan_extinction_2024}. This result adds weight to growing evidence---for example from dust polarization---for significant grain growth in protostellar envelopes. 
     
     \item 
     If the MIR emission lines trace lower electron density and temperature than the NIR lines, the inferred MIR extinction falls within the range of the existing values for dark clouds. In this case, there is no direct evidence for a change in the dust size distribution (i.e. grain growth) in protostellar envelopes beyond that in dense dark clouds. Instead, the observed MIR/NIR line ratios and excitation modeling suggest that the MIR and NIR lines originate from different regions along the same sightline within the embedded jet, each characterized by distinct physical conditions.

\end{enumerate}

\begin{acknowledgements}
This work was supported in part by an ALMA SOS award, STScI grant JWST-GO-02104.002-A, NSF grant AST-2307199, NASA ATP grant 80NSSC20K0533, and the Virginia Institute of Theoretical Astronomy.  This work is based on observations made with the NASA/ESA/CSA James Webb Space Telescope.  The data were obtained from the Mikulski Archive for Space Telescopes at the Space Telescope Science Institute, which is operated by the Association of Universities for Research in Astronomy, Inc., under NASA contract NAS 5-03127 for JWST. These observations are associated with GO program \#2104.
D.H. is supported by a Center for Informatics and Computation in Astronomy (CICA) grant and grant number 110J0353I9 from the Ministry of Education of Taiwan. D.H. also acknowledges support from the National Science and Technology Council, Taiwan (Grant NSTC111-2112-M-007-014-MY3, NSTC113-2639-M-A49-002-ASP, and NSTC113-2112-M-007-027).  A.C.G. acknowledges support from PRIN-MUR 2022 20228JPA3A “The path to star and planet formation in the JWST era (PATH)” funded by NextGeneration EU and by INAF-GoG 2022 “NIR-dark Accretion Outbursts in Massive Young stellar objects (NAOMY)” and Large Grant INAF 2022 “YSOs Outflows, Disks and Accretion: towards a global framework for the evolution of planet forming systems (YODA)”. D.R. acknowledges support by the European Research
Council advanced grant H2020-ER-2016-ADG-743029. We also thank Sophia Spiegel for contributing to the schematic shown in Figure 6, which enhances the clarity and impact of this work. 
\end{acknowledgements}

%
%
\bibliographystyle{aa}
\bibliography{jwst}

\begin{appendix}
\onecolumn
\section{Supplemental Material}
This appendix contains supplemental tables referenced throughout the main text. Table~\ref{tab:spectroscopic_data} lists atomic properties and critical densities for the [\ion{Fe}{II}] transitions analyzed in this study. Table~\ref{tab:integrated_intensities} provides the observed continuum-subtracted line intensities and ratios for each jet location. Table~\ref{tab:extinction_magnitude} summarizes the extinction values ($A_{1.644}$ and $A_V$) derived from various extinction curves. Table~\ref{table:electron_densities_a4D} presents the inferred electron densities based on corrected line ratios. 

\begin{table*}[h!]
\centering
\begin{tabular}{|c|cccc|cccc|c|c|c|}
\hline
$\lambda$ ($\mu$m) & Term$_k$ & $J_k$ & $g_k$ & Energy$_k$/k (K) & Term$_i$ & $J_i$ & $g_i$ & Energy$_i$/k (K) & $T_\text{ex}$ (K) & $A_{ki}$ (s$^{-1}$) & $n_\text{crit}$ (cm$^{-3}$) \\
\hline
25.988 & a6D & 7/2 & 8 & 554 & a6D & 9/2 & 10 & 0 & 554 & 2.05e-03 & 2.77e+04 \\
\hline
5.340 & a4F & 9/2 & 10 & 2694 & a6D & 9/2 & 10 & 0 & 2694 & 5.35e-05 & 1.33e+03 \\
17.936 & a4F & 7/2 & 8 & 3496 & a4F & 9/2 & 10 & 2694 & 802 & 6.01e-03 & 3.46e+04 \\
24.519 & a4F & 5/2 & 6 & 4083 & a4F & 7/2 & 8 & 3496 & 587 & 4.17e-03 & 2.22e+04 \\
\hline
1.257 & a4D & 7/2 & 8 & 11446 & a6D & 9/2 & 10 & 0 & 11446 & 5.18e-03 & 6.05e+04 \\
1.321 & a4D & 7/2 & 8 & 11446 & a6D & 7/2 & 8 & 554 & 10892 & 1.44e-03 & 6.05e+04 \\
1.372 & a4D & 7/2 & 8 & 11446 & a6D & 5/2 & 6 & 961 & 10485 & 8.93e-04 & 6.05e+04 \\
1.644 & a4D & 7/2 & 8 & 11446 & a4F & 9/2 & 10 & 2694 & 8752 & 5.73e-03 & 6.05e+04 \\
1.810 & a4D & 7/2 & 8 & 11446 & a4F & 7/2 & 8 & 3496 & 7950 & 1.25e-03 & 6.05e+04 \\
1.249 & a4D & 5/2 & 6 & 12074 & a6D & 7/2 & 8 & 554 & 11520 & 4.62e-04 & 5.11e+04 \\
1.295 & a4D & 5/2 & 6 & 12074 & a6D & 5/2 & 6 & 961 & 11113 & 2.23e-03 & 5.11e+04 \\
1.328 & a4D & 5/2 & 6 & 12074 & a6D & 3/2 & 4 & 1241 & 10833 & 1.30e-03 & 5.11e+04 \\
1.534 & a4D & 5/2 & 6 & 12074 & a4F & 9/2 & 10 & 2694 & 9380 & 3.00e-03 & 5.11e+04 \\
1.677 & a4D & 5/2 & 6 & 12074 & a4F & 7/2 & 8 & 3496 & 8578 & 2.38e-03 & 5.11e+04 \\
1.801 & a4D & 5/2 & 6 & 12074 & a4F & 5/2 & 6 & 4083 & 7991 & 1.72e-03 & 5.11e+04 \\
1.279 & a4D & 3/2 & 4 & 12489 & a6D & 3/2 & 4 & 1241 & 11248 & 2.80e-03 & 7.32e+04 \\
1.298 & a4D & 3/2 & 4 & 12489 & a6D & 1/2 & 2 & 1406 & 11083 & 1.23e-03 & 7.32e+04 \\
1.600 & a4D & 3/2 & 4 & 12489 & a4F & 7/2 & 8 & 3496 & 8993 & 4.03e-03 & 7.32e+04 \\
1.712 & a4D & 3/2 & 4 & 12489 & a4F & 5/2 & 6 & 4083 & 8406 & 1.13e-03 & 7.32e+04 \\
1.798 & a4D & 3/2 & 4 & 12489 & a4F & 3/2 & 4 & 4485 & 8004 & 2.00e-03 & 7.32e+04 \\
1.271 & a4D & 1/2 & 2 & 12729 & a6D & 1/2 & 2 & 1406 & 11323 & 3.82e-03 & 4.34e+04 \\
1.664 & a4D & 1/2 & 2 & 12729 & a4F & 5/2 & 6 & 4083 & 8645 & 4.56e-03 & 4.34e+04 \\
\hline
\end{tabular}

\caption{
Spectroscopic properties of [\ion{Fe}{II}] lines observed in TMC1A. Critical densities ($n_{\mathrm{crit}}$) for the Cloudy excitation model are calculated using Einstein A-coefficients and collisional strengths from \citet{verner1999numerical}, evaluated at $T = 7000$~K for consistency with previous shock studies \citep[e.g.,][]{koo_2016_shock_models}. The critical density is defined as $n_{\mathrm{crit}} = \sum A_{ki} / \sum C_{ki}$, where $A_{ki}$ is the spontaneous emission rate and $C_{ki} = \beta \Omega_{ki}(T) / (g_k \sqrt{T})$ is the collisional de-excitation rate. Here, $\Omega_{ki}(T)$ is the collision strength, $g_k$ the statistical weight of the upper level, and $\beta = 8.63 \times 10^{-6}$ a constant. The sums run over all radiative ($k > i$) and collisional ($k \neq i$) transitions from level $k$.
}

\label{tab:spectroscopic_data} 
\end{table*}

\begin{table*}[h!]
\centering
\begin{tabular}{|l |c  c  c c| c c c c| }
\hline
$\lambda$ ($\mu$m) & I (B) & I (F1) & I (F2) & I (F3) & R (B) & R (1) & R (2) & R (3) \\
\hline
25.988 & 1.85e-02 & 1.80e-02 & 1.28e-02 & 9.22e-03 & 1.058 & 2.558 & 7.668 & 4.471 \\
5.340 & 2.06e-02 & 1.39e-02 & 8.21e-03 & 1.45e-02 & 1.177 & 1.976 & 4.904 & 7.016 \\
17.936 & 3.26e-02 & 2.48e-02 & 1.04e-02 & 9.04e-03 & 1.867 & 3.517 & 6.225 & 4.384 \\
24.519 & 1.21e-02 & 6.36e-03 & 3.29e-03 & 1.60e-03 & 0.694 & 0.904 & 1.968 & 0.776 \\
\hline
1.257 & 3.14e-03 & 1.20e-03 & 3.17e-04 & 4.75e-04 & 0.180 & 0.171 & 0.189 & 0.231 \\
1.321 & 1.51e-03 & 6.20e-04 & 1.77e-04 & 2.13e-04 & 0.087 & 0.088 & 0.106 & 0.103 \\
1.372 & 1.18e-03 & 5.17e-04 & 1.26e-04 & 1.61e-04 & 0.068 & 0.073 & 0.075 & 0.078 \\
1.644 & 1.75e-02 & 7.04e-03 & 1.67e-03 & 2.06e-03 & 1.000 & 1.000 & 1.000 & 1.000 \\
1.810 & 5.53e-03 & 2.16e-03 & 5.28e-04 & 6.12e-04 & 0.317 & 0.307 & 0.315 & 0.297 \\
1.295 & --- & --- & --- & 8.00e-05 & --- & --- & --- & 0.039 \\
1.328 & 6.53e-04 & 2.55e-04 & 5.16e-05 & 6.51e-05 & 0.037 & 0.036 & 0.031 & 0.032 \\
1.534 & 3.47e-03 & 1.23e-03 & 2.34e-04 & 2.56e-04 & 0.199 & 0.174 & 0.140 & 0.124 \\
1.677 & 4.11e-03 & 1.44e-03 & 2.63e-04 & 2.52e-04 & 0.235 & 0.204 & 0.157 & 0.122 \\
1.801 & 4.09e-03 & 1.36e-03 & 2.53e-04 & 2.64e-04 & 0.234 & 0.193 & 0.151 & 0.128 \\
1.279 & 7.10e-04 & 2.57e-04 & 5.26e-05 & 4.42e-05 & 0.041 & 0.037 & 0.031 & 0.021 \\
1.298 & 2.89e-04 & 1.06e-04 & 1.47e-05 & 2.30e-05 & 0.017 & 0.015 & 0.009 & 0.011 \\
1.600 & 3.44e-03 & 1.17e-03 & 2.25e-04 & 2.10e-04 & 0.197 & 0.166 & 0.135 & 0.102 \\
1.712 & 1.20e-03 & 4.13e-04 & 8.20e-05 & 6.21e-05 & 0.069 & 0.059 & 0.049 & 0.030 \\
1.798 & 2.92e-03 & 9.44e-04 & 1.74e-04 & --- & 0.167 & 0.134 & 0.104 & --- \\
1.271 & 4.00e-04 & 1.59e-04 & 2.66e-05 & 2.27e-05 & 0.023 & 0.023 & 0.016 & 0.011 \\
1.664 & 2.23e-03 & 7.77e-04 & 1.49e-04 & 1.53e-04 & 0.128 & 0.110 & 0.089 & 0.074 \\
\hline
\end{tabular}
\caption{Continuum-subtracted integrated intensities of [\ion{Fe}{II}] lines observed at distinct jet locations. The first column is the rest-wavelength taken from NIST, then the 4 columns marked "I" represent the observed intensity at 4 distinct locations (in ergs/cm$^2$/s/sr) with an aperture of 0.7\arcsec. The columns marked "R" represent the ratio of that line with the 1.644 $\mu$m line. Missing data (---) is due to cases where artifacts (see Sec. \ref{Sec:Observations}) exist in the aperture (e.g.\ 1.295, 1.798 $\mu$m lines). 
}
\label{tab:integrated_intensities}
\end{table*}

\begin{table*}[h]
\centering
\begin{tabular}{|lcccc|}
\hline
& B & F1 & F2 & F3 \\
\hline
Observed $I_{1.257}/I_{1.644}$ & $0.18$ & $0.17$ & $0.19$ & $0.23$ \\
\hline
C89, R$_V$=5.5 & $3.78$ / $17.59$ & $3.88$ / $18.08$ & $3.68$ / $17.12$ & $3.28$ / $15.27$ \\
C89, R$_V$=3.1 & $3.78$ / $20.82$ & $3.88$ / $21.39$ & $3.68$ / $20.27$ & $3.28$ / $18.07$ \\
WD01, R$_V$=3.1 & $4.73$ / $18.12$ & $4.86$ / $18.63$ & $4.60$ / $17.64$ & $4.10$ / $15.73$ \\
WD01, R$_V$=5.5 & $4.79$ / $18.15$ & $4.92$ / $18.66$ & $4.66$ / $17.67$ & $4.16$ / $15.76$ \\
KP5 & $4.58$ / $18.12$ & $4.71$ / $18.62$ & $4.46$ / $17.64$ & $3.98$ / $15.73$ \\
Gordon, R$_V$=3.1 & $3.57$ / $22.22$ & $3.67$ / $22.84$ & $3.48$ / $21.63$ & $3.10$ / $19.29$ \\
Gordon, R$_V$=5.5 & $3.57$ / $23.36$ & $3.67$ / $24.01$ & $3.48$ / $22.74$ & $3.10$ / $20.28$ \\
\hline
Mean $A_{\lambda}$ $\pm$ Std Dev & $4.11 \pm 0.52$ & $4.23 \pm 0.53$ & $4.00 \pm 0.50$ & $3.57 \pm 0.45$ \\
Mean $A_V$ $\pm$ Std Dev & $19.77 \pm 2.17$ & $20.32 \pm 2.23$ & $19.25 \pm 2.11$ & $17.16 \pm 1.88$ \\
\hline

\end{tabular}
\caption{
Observed [\ion{Fe}{II}] 1.257/1.644 intensity ratios and the corresponding extinction values ($A_{1.644}$ and $A_V$) derived toward each jet location analyzed in this study (see Fig.~\ref{fig:iron_detections_subset}), calculated using various extinction curves (see caption of Fig.~\ref{fig:a4dj72_extinction}) and an intrinsic ratio of 1.18 taken from the atomic dataset used in the Cloudy model (see Sect. \ref{Sec:Fe2_model}). The top row lists the observed line ratios used to infer extinction, followed by extinction values derived from each extinction law, formatted as $A_{1.644}$ / $A_V$. The bottom two rows show the mean and standard deviation of $A_{1.644}$ and $A_V$ across all extinction laws for each location. The mean $A_{1.644}$ value is used to normalize the extinction differences $\Delta A_{\lambda}$, yielding $\beta_{\lambda} = A_{\lambda} / A_{1.644}$ in Section~\ref{Sec:4_Determine_Differences_in_Extinction}. The $A_V$ values are included to illustrate that differences in $R_V$ across extinction laws---specifically where both $R_V = 3.1$ and $R_V = 5.5$ are considered---have a smaller effect on near-IR extinction ($A_{1.644}$) compared to optical extinction ($A_V$). 
}

\label{tab:extinction_magnitude}
\end{table*}

\begin{table*}[h!]
\centering

\begin{tabular}{|c|c|c|c|c|c|c|c|c|}
\hline
$\lambda$ &  \multicolumn{4}{c|}{$\bar{R}_{corrected}$} & \multicolumn{4}{c|}{$\bar{n}_e$} \\
\hline
 & B & F1 & F2 & F3 & B & F1 & F2 & F3 \\
\hline
1.295 & 0.338 & 0.720 & 1.407 & 0.161 & --- & ---& --- & $0.8_{0.5}^{1.5}$ \\
1.328 & 0.158 & 0.159 & 0.125 & 0.110 & $4.7_{2.0}^{90.0}$ & $5.0_{2.0}^{90.0}$ & $2.0_{1.0}^{3.5}$ & $1.4_{0.8}^{2.0}$ \\
1.534 & 0.301 & 0.268 & 0.210 & 0.178 & $3.7_{2.0}^{20.0}$ & $2.4_{1.5}^{5.0}$ & $1.1_{0.7}^{2.0}$ & $0.8_{0.5}^{1.0}$ \\
1.677 & 0.209 & 0.181 & 0.140 & 0.110 & $3.1_{1.5}^{8.5}$ & $2.0_{1.0}^{3.5}$ & $0.9_{0.6}^{1.5}$ & $0.5_{0.3}^{0.8}$ \\
1.801 & 0.143 & 0.116 & 0.093 & 0.083 & $3.3_{1.5}^{10.0}$ & $1.7_{0.9}^{3.0}$ & $0.9_{0.5}^{1.5}$ & $0.7_{0.4}^{1.0}$ \\
1.279 & 0.230 & 0.217 & 0.170 & 0.097 & $6.8_{3.5}^{100.0}$ & $5.1_{2.5}^{90.0}$ & $2.4_{1.5}^{3.5}$ & $0.7_{0.5}^{1.0}$ \\
1.298 & 0.084 & 0.080 & 0.043 & 0.046 & $3.3_{2.0}^{9.5}$ & $2.9_{1.5}^{6.5}$ & $0.8_{0.5}^{1.0}$ & $0.9_{0.6}^{1.0}$ \\
1.600 & 0.229 & 0.194 & 0.156 & 0.116 & $3.6_{2.0}^{10.0}$ & $2.4_{1.5}^{3.5}$ & $1.5_{0.9}^{2.0}$ & $0.8_{0.6}^{1.0}$ \\
1.712 & 0.055 & 0.047 & 0.039 & 0.025 & $2.8_{1.5}^{5.0}$ & $2.0_{1.0}^{3.0}$ & $1.4_{0.9}^{2.0}$ & $0.6_{0.4}^{0.8}$ \\
1.798 & 0.102 & 0.081 & 0.065 & 0.116 & $3.7_{2.0}^{20.0}$ & $2.0_{1.0}^{3.0}$ & $1.2_{0.8}^{2.0}$ & --- \\
1.271 & 0.137 & 0.142 & 0.091 & 0.052 & $4.0_{2.5}^{20.0}$ & $4.6_{2.5}^{85.0}$ & $1.5_{1.0}^{2.0}$ & $0.6_{0.4}^{0.8}$ \\
1.664 & 0.119 & 0.103 & 0.083 & 0.070 & $3.5_{2.0}^{9.5}$ & $2.3_{1.5}^{3.5}$ & $1.5_{1.0}^{2.0}$ & $1.0_{0.7}^{1.5}$ \\

\hline
\multicolumn{5}{|c|}{Average n$_e$} & 3.6e+04 ± 1e+04 & 2.5e+04 ± 0.6e+04 & 1.5 ± 5e+03 & 8.0e+03 ± 3e+03 \\
\hline

\end{tabular}

\caption{Estimated electron densities calculated using the 'a4D' line intensity ratios relative to the 1.644 $\mu$m line. The extinction magnitude is calculated for each extinction curve, and each curve is used to correct for extinction at each line's wavelength. This process yields a range of corrected line ratios, with deviations on the order of $10^{-3}$ due to differences among the extinction curves. The reported standard deviation for the electron density accounts for both the variation among corrected line ratios for each extinction curve and the variation in electron densities matched to the model at each temperature ($T = 5\times 10^3 - 2 \times 10^4$ K). The final row shows the weighted average electron density and its weighted standard deviation. The results reveal a consistent trend of decreasing electron density with increasing distance from the jet.}
\label{table:electron_densities_a4D}
\end{table*}

\end{appendix}
\end{document}